\documentclass[sigconf]{acmart} 

\usepackage{enumitem}
\usepackage{subcaption}
\usepackage{graphicx}
\usepackage{balance}
\usepackage{amsmath}
\usepackage{color}
\usepackage{xcolor}
\usepackage{booktabs}
\usepackage{subcaption}

\usepackage{makecell}

\AtBeginDocument{%
  \providecommand\BibTeX{{%
    \normalfont B\kern-0.5em{\scshape i\kern-0.25em b}\kern-0.8em\TeX}}}

\newcommand{\RA}[1]{\href{https://www.runningahead.com/logs/#1/workouts}{\textcolor{blue}{\texttt{#1}}}}
\newcommand{\strava}[1]{\href{https://www.strava.com/activities/#1}{\textcolor{blue}{\texttt{#1}}}}

\copyrightyear{2025}
\acmYear{2025}
\setcopyright{rightsretained}
\acmConference[CHIGreece 2025]{CHIGreece 2025: 3rd International Conference of the ACM Greek SIGCHI Chapter}{September 24--26, 2025}{Hermoupolis, Syros, Greece}
\acmBooktitle{CHIGreece 2025: 3rd International Conference of the ACM Greek SIGCHI Chapter (CHIGreece 2025), September 24--26, 2025, Hermoupolis, Syros, Greece}
\acmPrice{}
\acmDOI{10.1145/3749012.3749046}
\acmISBN{979-8-4007-1561-7/25/09}

\begin{document}

\title{A Longitudinal Evaluation of Heart Rate Efficiency for Amateur Runners}

\author{Evgeny V. Votyakov}
\orcid{0000-0002-6345-4726}
\affiliation{
  \institution{CYENS Centre of Excellence}
  \city{Nicosia}
  \country{Cyprus}}
\email{e.votyakov@cyens.org.cy}

\author{Marios Constantinides}
\orcid{0000-0003-1454-0641}
\affiliation{
  \institution{CYENS Centre of Excellence \& University of Cyprus}
  \city{Nicosia}
  \country{Cyprus}}
\email{marios.constantinides@cyens.org.cy}

\author{Fotis Liarokapis}
\orcid{0000-0003-3617-2261}
\affiliation{
  \institution{CYENS Centre of Excellence}
  \city{Nicosia}
  \country{Cyprus}}
\email{f.liarokapis@cyens.org.cy}

\renewcommand{\shortauthors}{Votyakov et al.}

\begin{abstract}
Amateur runners are increasingly using wearable devices to track their training, and often do so through simple metrics such as heart rate and pace. However, these metrics are typically analyzed in isolation and lack the explainability needed for long-term self-monitoring. In this paper, we first present Fitplotter, which is a client-side web application designed for the visualization and analysis of data associated with fitness and activity tracking devices. Next, we revisited and formalized Heart Rate Efficiency (HRE), defined as the product of pace and heart rate, as a practical and explainable metric to track aerobic fitness in everyday running. Drawing on more than a decade of training data from one athlete, and supplemented by publicly available logs from twelve runners, we showed that HRE provides more stable and meaningful feedback on aerobic development than heart rate or pace alone. We showed that HRE correlates with training volume, reflects seasonal progress, and remains stable during long runs in well-trained individuals. We also discuss how HRE can support everyday training decisions, improve the user experience in fitness tracking, and serve as an explainable metric to proprietary ones of commercial platforms. Our findings have implications for designing user-centered fitness tools that empower amateur athletes to understand and manage their own performance data.
\end{abstract}


\maketitle

\section{Introduction}
\label{sec:introduction}
Recreational running has become increasingly data-driven, with amateur athletes routinely using wearable technologies (e.g., smartwatches) to track their training progress~\cite{rooksby2014personal, li2010stage}. These technologies offer easy access to physiological and performance metrics such as pace, distance, and heart rate (HR), which users often rely on to guide their pacing strategies and training load adjustments. However, despite the ubiquity of these technologies, their ability to support reflective self-monitoring and long-term training insights remains limited. Prior Human-Computer Interaction (HCI) research has emphasized the importance of designing personal informatics tools that support meaningful feedback and self-awareness, particularly in health and fitness contexts~\cite{epstein2015lived, choe2014understanding}. Yet, most consumer-grade fitness platforms prioritize surface-level metrics over explainability, leaving users with raw data rather than actionable understanding. While similar heuristics exist in endurance sports communities (e.g., Relative Running Economy used in RUNALYZE~\cite{runalyze}), they often require additional parameters such as resting HR, and are rarely subject to empirical evaluation.

Metrics such as heart rate and pace are also accessible, but they are typically interpreted in isolation and offer only partial views of a runner's physiological state. Additionally, commercial fitness platforms often rely on opaque algorithms to generate performance summaries (e.g., ``training readiness'' indicators) without disclosing how these metrics are computed. This lack of explainability can undermine user trust, reduce perceived usefulness, and limit the user's ability to critically engage with their own data~\cite{kay2012lullaby, choe2014understanding}. As a result, amateur runners are left with data that may be inconsistent, difficult to interpret over time, and disconnected from broader training goals. HCI scholars have long cautioned against opaque or overly abstract feedback in self-tracking systems, which can hinder long-term engagement and reduce perceived value~\cite{beer2017quantified, nunes2015self}. To this end, there is a pressing need for metrics that are empirically grounded, transparent, and meaningful over extended periods of self-directed training.

In this work, we revisited and formalized Heart Rate Efficiency (HRE) as a practical and explainable metric for monitoring aerobic fitness in amateur running. Drawing on over a decade of personal training data from one athlete and supplemented with public logs from 12 runners, we analyzed how HRE evolves over time, correlates with training volume, and remains stable during long steady-state runs. We argue that HRE can serve as a transparent proxy for aerobic fitness by offering a user-centered alternative to proprietary black-box metrics. Our contributions are two-fold: \emph{1)} We developed FitPlotter, a client-side web application for visualizing and analyzing fitness and activity tracking data, and formalized HRE as a user-friendly and explainable metric for tracking aerobic fitness (\S\ref{sec:methods}); and \emph{2)} Using FitPlotter, we analyzed a longitudinal dataset spanning over ten years of training from one athlete along with supplementary dataset of 12 additional runners, and demonstrated how HRE reflects aerobic development more reliably than heart rate or pace alone (\S\ref{sec:results}).

While our study draws on a longitudinal dataset, it remains exploratory in nature. Drawing from this analysis, we discuss the potential of HRE to support everyday training, and propose how fitness tracking platforms can integrate such metrics to enhance explainability and user trust (\S\ref{sec:discussion}). By making HRE accessible and empirically grounded, we aim to contribute to the design of more transparent and meaningful self-tracking tools for recreational athletes.

\section{Related Work}
\label{sec:related}
We surveyed various lines of research that our work draws upon and grouped them into four main
areas: \emph{i)} empirical and theoretical pacing models; \emph{ii)} self-tracking and fitness metrics in HCI; \emph{iii)} understanding aerobic fitness; and \emph{iv)} designing transparent and empowering fitness tools.
\smallskip

\noindent\textbf{Empirical and Theoretical Pacing Models.} Endurance running is shaped by a complex interplay of physiological, psychological, and environmental factors. Classic sports science literature highlights ${VO}_{2}$ max, lactate threshold, and running economy as key physiological indicators of endurance capacity~\cite{Daniels1979,Billat2003,Faude2009}. Building on these foundations, empirical pacing models (e.g., Riegel's rule~\cite{Riegel1981}), have been widely used to forecast performance based on training history and race distance, and generate personalized pacing plans~\cite{Altini2016,Cvetkovic2020}. These models account for training load, weather, elevation, and runner-specific characteristics to improve real-world applicability~\cite{Ely2007,Fister2015}. Other approaches emphasize the psychological dimension of pacing. Foster et al. \cite{Foster1994} describe a teleoanticipatory mechanism where perceived effort influences pace, while Stevinson and Biddle \cite{Stevinson1998} link mental strategies to the experience of ``hitting the wall''. Machine learning systems further tailor predictions by leveraging large race datasets and case-based reasoning~\cite{Berndsen2019,Smyth2017} to provide personalized training guidance~\cite{feely2024pseudo,Feely2024}; such systems were found to augment expert coach judgment~\cite{Rutjes2024}.
\smallskip

\noindent\textbf{Self-Tracking and Fitness Metrics in HCI.} Fitness tracking has been a longstanding interest in HCI, personal informatics, and quantified-self movements~\cite{rooksby2014personal, li2010stage, park2020wellbeat} to support self-awareness and behavioral change~\cite{park2023social}. While physiological data such as heart rate, steps, and sleep can be easily captured, prior work has highlighted their limitations in supporting meaningful reflection and sustained engagement~\cite{beer2017quantified}. This is particularly true in fitness contexts, where users often receive abstract or proprietary metrics (e.g., ``training readiness'' indicators) without understanding how they are derived or how to act on them~\cite{kay2012lullaby, choe2014understanding}.

To understand how trackers engage users, the Tracker Goal Evolution Model outlined how tracking goals develop from intrinsic user needs into qualitative and ultimately quantitative goals suitable for use with fitness trackers~\cite{Niess2018}. Moreover, HCI researchers have advocated for more interpretable and user-centered representations of physiological data. Work by Epstein et al.~\cite{epstein2015lived} and Choe et al.~\cite{choe2014understanding} emphasizes the need for systems that support long-term tracking, contextual interpretation, and user trust. 
\smallskip

\noindent\textbf{Understanding Aerobic Fitness.} In sports science, aerobic fitness is commonly assessed through metrics such as the lactate threshold or ${VO}_{2}$ max, which are often measured under controlled laboratory conditions~\cite{joyner2008endurance}. While accurate, these assessments are impractical for most amateur runners, who depend instead on heuristics such as pace and average heart rate. In response, informal proxies like the Maximum Aerobic Function (MAF) test and ``pulse cost'' are used among running communities. These use consistent pacing or heart rate to evaluate training progress, and offer an alternative to lab testing~\cite{maffetone2010big}. Commercial fitness platforms have introduced simplified heuristics to approximate training load or aerobic fitness. However, users often find them unreliable or difficult to interpret over time due to their sensitivity to contextual factors like weather, terrain, and fatigue~\cite{choe2014understanding, gouveia2015we}. Despite their limitations, they reveal an important need: \emph{the desire among runners to make sense of their fitness data without requiring complex physiological expertise}. 

Other metrics commonly discussed in exercise science include Running Economy (RE)~\cite{Saunders2004}, which is the oxygen cost at a submaximal pace, and the Cardiac Drift~\cite{Coyle2001}, which captures the gradual increase in heart rate during steady-state exercise. These metrics typically require controlled environments for accurate measurement. Some commercial platforms offer alternatives such as the Efficiency Index (running speed divided by oxygen consumption). While commercial devices offer alternative performance metrics, these are rarely validated across long-term or broad-spectrum datasets. Many rely on proprietary correlations with ${VO}_{2}$ max, typically measured in controlled lab environments using gas analysis~\cite{Smyth2022}. For example, the Relative Running Economy (RRE) used by Runalyze requires resting heart rate, which itself varies significantly over a training cycle. By contrast, HRE uses only pace and HR (metrics already familiar to runners) and is easier to interpret without calibration. While well-calibrated systems may converge on similar trends, HRE's simplicity and explainability make it more practical for everyday self-monitoring.
\smallskip

\noindent\textbf{Designing Transparent and Empowering Fitness Tools.} One recurring theme in HCI is the need for self-tracking that go beyond passive data collection to actively empower users~\cite{nunes2015self, suh2014babysteps}. This includes not only presenting understandable feedback but also scaffolding long-term engagement through personalization, reflection, and support for routine integration~\cite{baumer2015reflective}. In the context of fitness tracking, such systems should align with users' goals (e.g., marathon readiness), acknowledge physical and psychological diversity, and avoid over-reliance on proprietary black-box algorithms. Studies show that users are more likely to trust and continue using systems when they understand how feedback is calculated and can verify improvements over time~\cite{gouveia2015we}. However, most existing platforms prioritize breadth of metrics over interpretability, often resulting in overload or disengagement. 
\smallskip

\noindent\textbf{Research Gaps.} While prior HCI research has explored the design of self-tracking systems and the use of wearables in fitness contexts~\cite{rooksby2014personal, li2010stage, park2020wellbeat, nunes2015self}, few studies have examined longitudinal metrics that are both explainable and practically useful for amateur athletes. Existing metrics are either difficult to interpret (e.g., ${VO}_{2}$ max scores) or too volatile to track meaningful changes over time (e.g., heart rate alone). There is a lack of empirical, user-centered research on simplified yet robust heuristics for monitoring aerobic development. In this paper, we address that gap by revisiting Heart Rate Efficiency (HRE), analyzing over ten years of training data, and offering design insights for how such metrics could be integrated into everyday fitness tools.
\begin{table}[t!]
  \centering
  \caption{Athletes labels for aggregated HRE and activity labels for HRE dynamics. For  RA:  \url{https://www.runningahead.com/logs/RA-id}, for strava: \url{https://www.strava.com/activities/act-id}. Depending on the pdf viewer, click on *-id opens the corresponding URL.}
  \label{tab:public_athletes}
  \scalebox{0.55}{
  \begin{tabular}{l r || l r || l r}
    \toprule
    \textbf{Label} & \textbf{RA-id} & 
    \textbf{Label} &  \textbf{Strava act-id} &
    \textbf{Label} &  \textbf{Strava act-id} \\
    \midrule
    A & \RA{4e4ce3edd52e49049071460b75d1eca8} & ALM2018 & \strava{1971185275} &  2016 Prague & \strava{569343111} \\
    B & \RA{587be746fe014316945cee19169b7a44} & BLM2018 & \strava{1972257898}  &  2017 Berlin & \strava{1199024727} \\
    C & \RA{636d955284d14509b91fc61b02781d57} & CLM2018 & \strava{1971563210} &  2018 Chicago & \strava{1891059860} \\
    D & \RA{d8e690f304014a41bdbef78bc84a1cfc} & DLM2017 & \strava{1281673153}   &  2019 Boston & \strava{2292241786} \\
    E & \RA{a2da1747a10f4462bfa9422f915ab312} & ELM2018 & \strava{1971392905} &  2019 Valencia & \strava{2903579996} \\
    F & \RA{17d3bd8eb29d439394ac4ba41ad22de6} & FLM2018 & \strava{1971476020}  &  2020 Bedford & \strava{4210796866} \\
    G & \RA{920364479057491fac580b21d5618698} & GLM2017 & \strava{1281741250} & 2021 London & \strava{6057170501} \\
    H & \RA{822357b8276f4d699fe9d18c757e087e} & & & 2022 SanSebastian & \strava{8174923272}  \\
    I & \RA{18f66f47702e4b1aa74249d6ca3503de} & & & 2023 LaRochelle  & \strava{10283952164} \\
    J & \RA{6afe15bf9502462bb8ad33319e14db7a} & & & & \\
    K & \RA{214a42f5e65141d7a9c737d0daebc7c6} & & & & \\
    L & \RA{6d5a0bc0d3a94017b98b500af5ed1450} & & & & \\
    M & \RA{c9850b9f68d34980949cc544edec07ed} & & & & \\
\bottomrule
  \end{tabular}
  }
\end{table}

\section{Methods}
\label{sec:methods}

\noindent\textbf{Fitplotter: A Tool for Fitness Data Visualization.} We developed Fitplotter,\footnote{ \url{https://anonymous.4open.science/r/fitplotter-388E/README.md} and online: \url{https://anonymous.4open.science/w/fitplotter-388E/}} which is a client-side web application to support the visualization and analysis of fitness data from common training devices (e.g., Garmin, Suunto). It enables users to inspect .FIT and .TKL files without relying on cloud services or proprietary platforms by allowing training data to be loaded and analyzed entirely within the browser. Once files are loaded, Fitplotter displays interactive charts (using CanvasJS) for exploring parameters such as heart rate, pace, and HRE across sessions. Users can zoom into specific ranges, toggle data series, and overlay multiple datasets to support comparative analysis~ \footnote{It is absent in Garmin and Strava software.}. Fitplotter also integrates a synchronized map (via Leaflet) that allows users to link specific chart segments with their corresponding GPS coordinates\footnotemark[\value{footnote}]. This is particularly helpful for analyzing terrain effects or event-specific variations in HRE. To enhance our tool's usability, we included a lap-based segmentation for interval or fatigue analysis, data smoothing to mitigate sensor noise, and real-time breathing rate estimation using Heart Rate Variability (HRV).

The development of Fitplotter was motivated by practical needs encountered during long-term self-tracking and grounded on previous literature~\cite{choe2014understanding,epstein2015lived}. These include the difficulty of accessing and interpreting raw training data across platforms, and the lack of tools to visualize trends in a format suitable for personal reflection. Rather than optimizing for automation or recommendations, Fitplotter focuses on data transparency and usability by allowing runners to explore their own .FIT and .TKL files locally without uploading to third-party servers. All visualizations and computations were selected form *.FIT logs using Fitplotter and then plotted with standard Python means. In line with open-science principles, we made our tool freely available under the MIT license; this also aligns with open-science values of promoting transparency, reproducibility, and community reuse. To support more advanced workflows, we also developed two optional companion tools: Fitalyser, a batch-processing interface for aggregating statistics across multiple sessions, and GChandler, which automates the download of activity files from Garmin Connect. Together, these tools provide a modular platform for researchers, developers, and athletes who regularly analyze fitness data at scale.

\noindent\textbf{Data Sources.} This study draws upon three primary data sources (Table~\ref{tab:public_athletes}). First, a longitudinal self-tracking dataset compiled by the first author (athlete A) over the course of more than a decade (2011--2024) of personal training. This dataset comprises over 4,000 sessions, including GPS, elevation, and timestamped information on average HR, pace (min/km), distance, and event type, covering marathon, ultra-marathon, mountain trail, and triathlon training activities. Second, a publicly available training logs from 12 endurance anonymous runners, each with several hundred recorded sessions. The criteria for athlete selection and data curation are described below. Third, sixteen activity records from recreational and semi-professional runners, used for case-specific comparisons.

Supplementary data were obtained from two publicly accessible fitness tracking platforms: RunningAHEAD (RA)\footnote{\url{https://www.runningahead.com/}} and Strava\footnote{\url{https://www.strava.com/}}. RA hosts an active online community and organizes annual distance-based challenges (e.g., 20XX Miles in 20XX) where participants aim to complete a number of miles equal to the calendar year. Archived editions of these challenges\footnote{\url{https://www.runningahead.com/distance_game/2016}} enable the manual identification of runners with consistently high training volumes over multiple years.
 
Athlete selection for the RA dataset was based on the following criteria: (a) public availability of training logs, (b) inclusion of heart rate data, and (c) a minimum of one year of consistent training activity. Due to inherent platform limitations, it was not possible to determine the total number of challenge participants or how many met these criteria. Furthermore, accessing private data without user consent would violate RA's terms of service. All data collection was performed manually, and all personal identifiers (e.g., names, ages, genders) were anonymized. Only activity-level metadata accessible via RA-ids were retained for analysis. For the HRE dynamics analysis, marathon-specific act-ids were retrieved from Strava’s event-based aggregation feature, which clusters activity records from athletes participating in the same event. Selection criteria included publicly shared activity records with visible HR and pace data.
\smallskip

\noindent\textbf{Ethical Considerations.} The first author collected their data for personal use. Marathon data from a single athlete that participated in the 2016 Prague Marathon to the 2023 La Rochelle Marathon were included with explicit consent. For the rest of the reported data in this study, we used publicly shared running logs that were freely accessible.
\smallskip

\begin{table*}[t!]
    \centering
    \begin{minipage}{0.4\textwidth}
        \centering
        \small
        \caption{Example of athlete's A data used in analysis.}
        \begin{tabular}{rccccc}
            \toprule
            \makecell{Date} & \makecell{Distance \\ (km)} & \makecell{Pace \\ (min/km)} & \makecell{HR\\(bpm)} & \makecell{Pace \\ (decimal)} & \makecell{HRE\\(bpkm)} \\
            \midrule
            8/27/2018 & 15.7 & 5:25 & 140 & 5.42 & 758 \\
            ...  & ... & ... & ... & .. & ... \\
            8/9/2018  & 12.8 & 5:48 & 143 & 5.80 & 829 \\
            ...  & ... & ... & ... & .. & ... \\
            8/2/2018  & 11.0 & 5:50 & 147 & 5.83 & 857 \\
            ...  & ... & ... & ... & .. & ... \\
            7/30/2018 & 10.5 & 6:04 & 144 & 6.07 & 874 \\
            \bottomrule
        \end{tabular}
        
        \label{tab:HRE_august}
    \end{minipage}%
    \hfill
    \begin{minipage}{0.48\textwidth}
        \centering
        \includegraphics[width=0.7\textwidth]{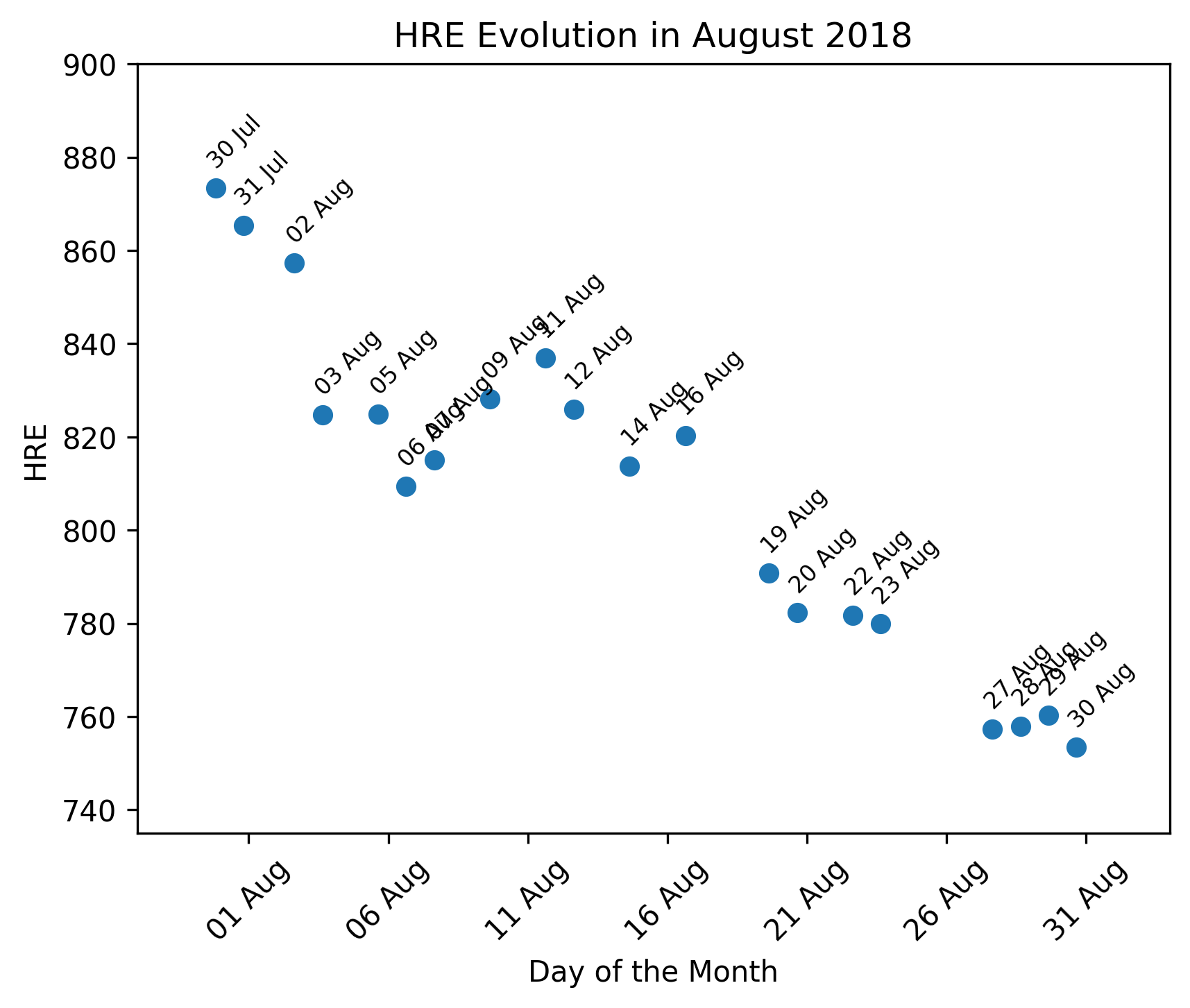}
        \captionof{figure}{Athlete's A HRE evolution.}
        \label{fig:HRE_august}
    \end{minipage}
\end{table*}

\noindent\textbf{Heart Rate Efficiency (HRE).} It is a  product of heart rate (in beats per minute) and running pace (in minutes per kilometer). It is of an unit of beats per kilometer: $ \text{HRE} = \text{HR} \times \text{Pace}$, i.e. the number of heartbeats required to cover one kilometer, integrating both cardiovascular effort and movement efficiency.  Unlike laboratory-grade indicators such as ${VO}_{2}$ max, HRE can be simply and fast calculated using data available to any runner with a modern GPS watch.

HRE  is a metric that is less influenced by the athlete's instantaneous effort compared to pace or HR. If one aims to achieve a faster pace, it comes proportionally at the cost of a higher HR. Conversely, if one tries to maintain a lower HR, the pace naturally decreases. In both cases, HRE tends to be the same in the sustainable range when the athlete feels comfortably. Therefore, HRE is not directly subject to the athlete's conscious control, thus making it a more objective indicator of current performance. Over the long term, average HRE improves, if the athlete trains correctly.
\smallskip

\noindent\textbf{Analysis.} Our analysis focused on two complementary aspects of Heart Rate Efficiency (HRE), grounded in longitudinal training data from Athlete A. First, we examined how HRE evolves over time in response to structured aerobic training. To ensure comparability across sessions, we identified runs performed under similar conditions—same route, distance, and weather. One such example, presented in Table~\ref{tab:HRE_august}, captures a series of aerobic sessions during August 2018 following a period of injury. The data show a progressive improvement in HRE, approximately 10–15\% over the course of the month, thus reflecting recovery of aerobic capacity. Figure~\ref{fig:HRE_august}, visualizes this trend, illustrating HRE's sensitivity to incremental gains in efficiency even over short periods. 
Second, we analyzed intra-run HRE dynamics during long-distance sessions lasting over 90 minutes. These sessions revealed that well-trained runners tend to maintain a stable HRE throughout flat routes, whereas less trained runners often exhibit degradation in HRE after 30–40 minutes, indicative of fatigue. On hilly terrain, HRE fluctuations aligned with the route's elevation profile, with elevated values on uphill segments and lower values during descents. These patterns support the utility of HRE as both a longitudinal indicator of aerobic development and a real-time signal for assessing physiological pacing and endurance limits.
\section{Results}
\label{sec:results}

\begin{table}[t!]
  \centering
  \caption{Athlete A: Yearly training summary and Heart Rate Efficiency (HRE) statistics (2011–2024).}
  \label{tab:yearly_summary}
  \scalebox{0.83}{
  \begin{tabular}{lcccccc}
    \toprule
    \textbf{\makecell{Year}} & \textbf{\makecell{Avg. Monthly\\Distance (km)}} & \textbf{\makecell{Avg. Pace\\(min/km)}} & \textbf{\makecell{Avg. HR\\ (bpm)}} & \textbf{\makecell{Avg. HRE\\(bpkm)}} & \textbf{\makecell{Min HRE\\(bpkm)}} \\
    \midrule
    2011 & 186 & 6:01 & 148 & 902 & 740 \\
    2012 & 292 & 6:07 & 143 & 889 & 720 \\
    2013 & 283 & 5:18 & 146 & 776 & 710 \\
    2014 & 297 & 5:20 & 144 & 770 & 700 \\
    2015 & 334 & 5:50 & 141 & 822 & 698 \\
    2016 & 310 & 5:49 & 138 & 798 & 685 \\
    2017 & 280 & 5:55 & 133 & 785 & 695 \\
    2018 & 266 & 6:19 & 130 & 798 & 690 \\
    2019 & 207 & 6:04 & 134 & 798 & 703 \\
    2020 & 295 & 5:26 & 138 & 750 & 690 \\
    2021 & 221 & 5:37 & 135 & 761 & 707 \\
    2022 & 167 & 5:46 & 138 & 795 & 735 \\
    2023 & 138 & 5:57 & 138 & 825 & 742 \\
    2024 & 105 & 5:46 & 139 & 805 & 754 \\
    \bottomrule
  \end{tabular}
  }
\end{table}

\begin{figure*}[t!]
    \centering

    \begin{subfigure}[b]{0.22\textwidth}
        \includegraphics[width=\textwidth]{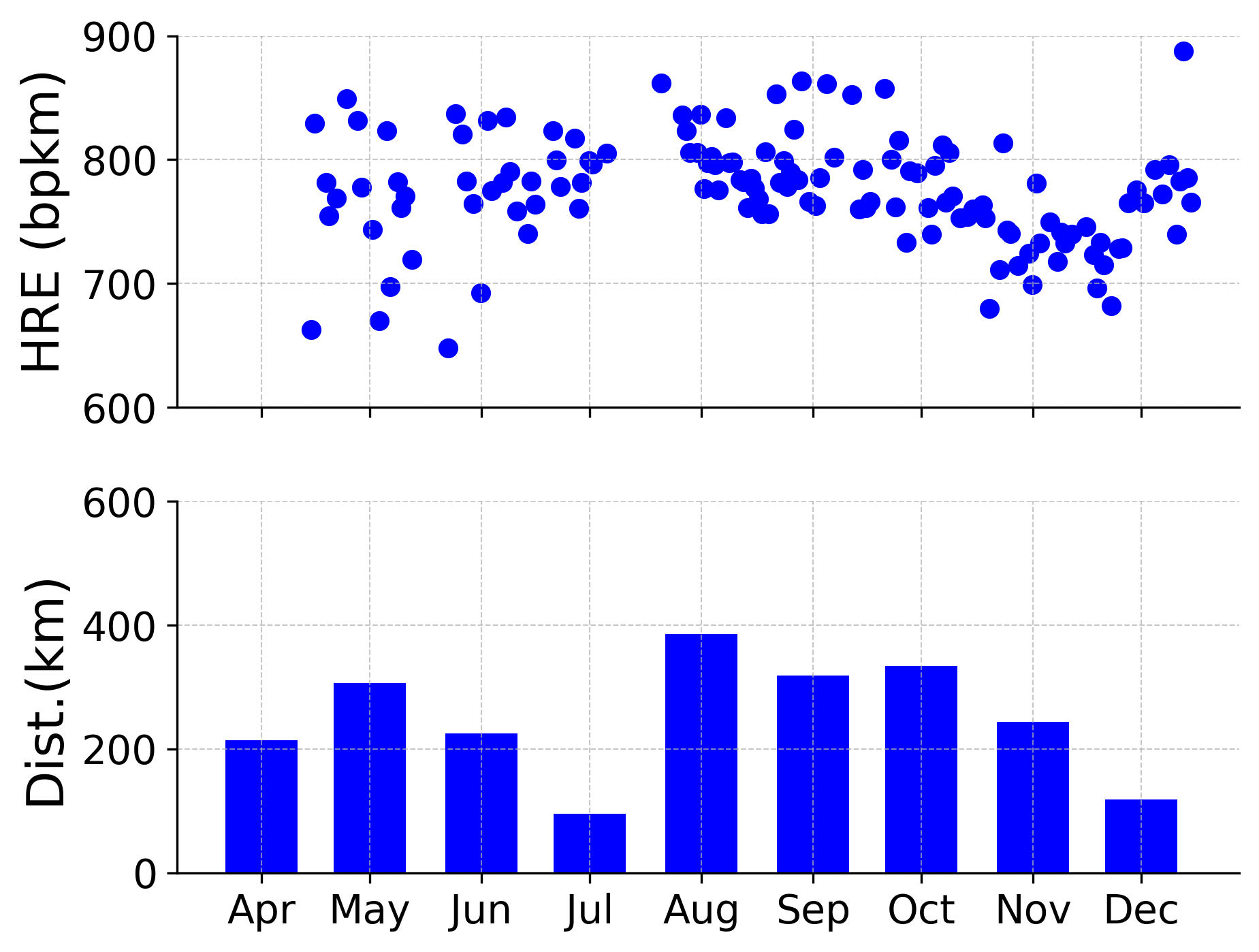}
        \caption{2011}
    \end{subfigure}
    \hfill
    \begin{subfigure}[b]{0.22\textwidth}
        \includegraphics[width=\textwidth]{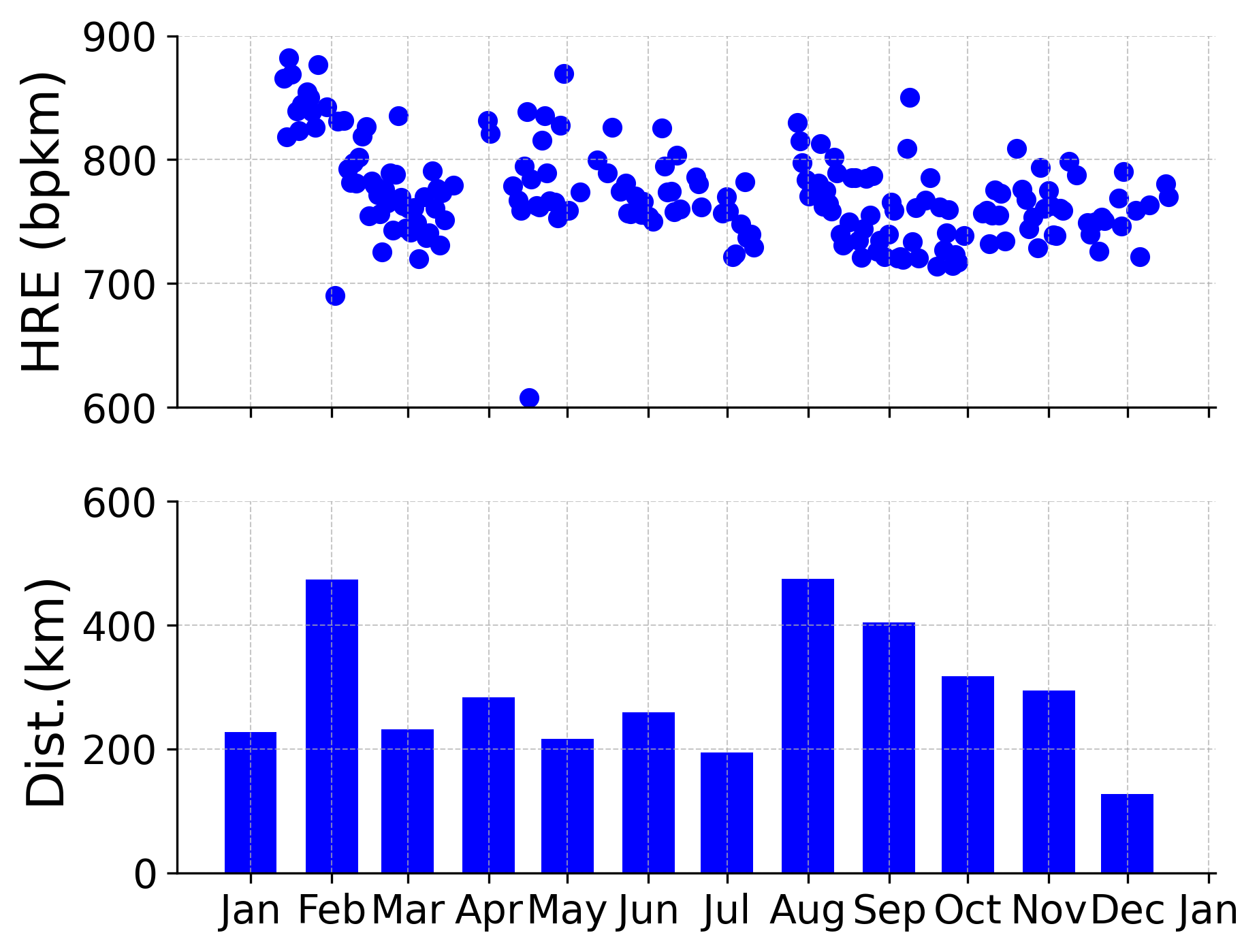}
        \caption{2012}
    \end{subfigure}
    \hfill
    \begin{subfigure}[b]{0.22\textwidth}
        \includegraphics[width=\textwidth]{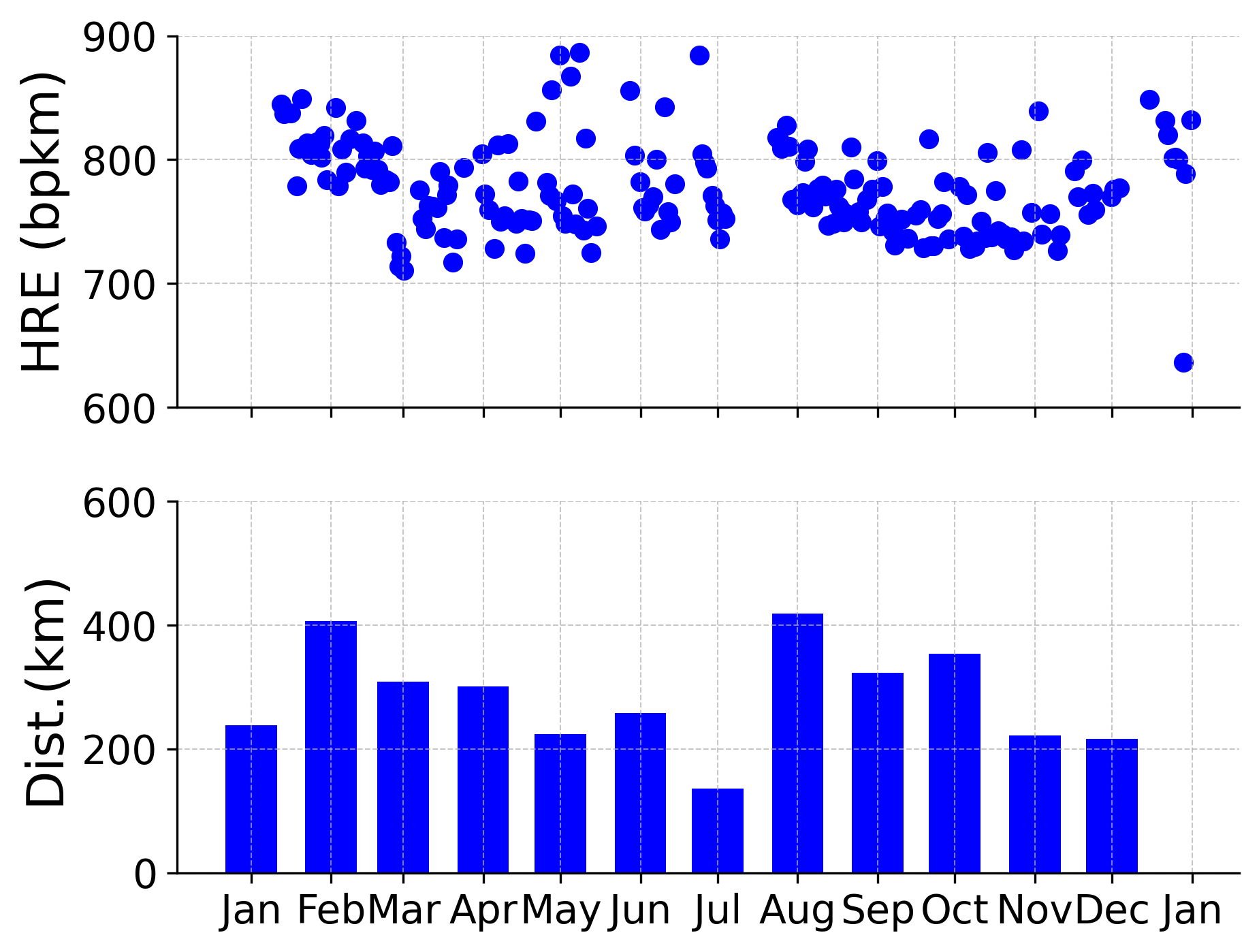}
        \caption{2013}
    \end{subfigure}
    \hfill
    \begin{subfigure}[b]{0.22\textwidth}
        \includegraphics[width=\textwidth]{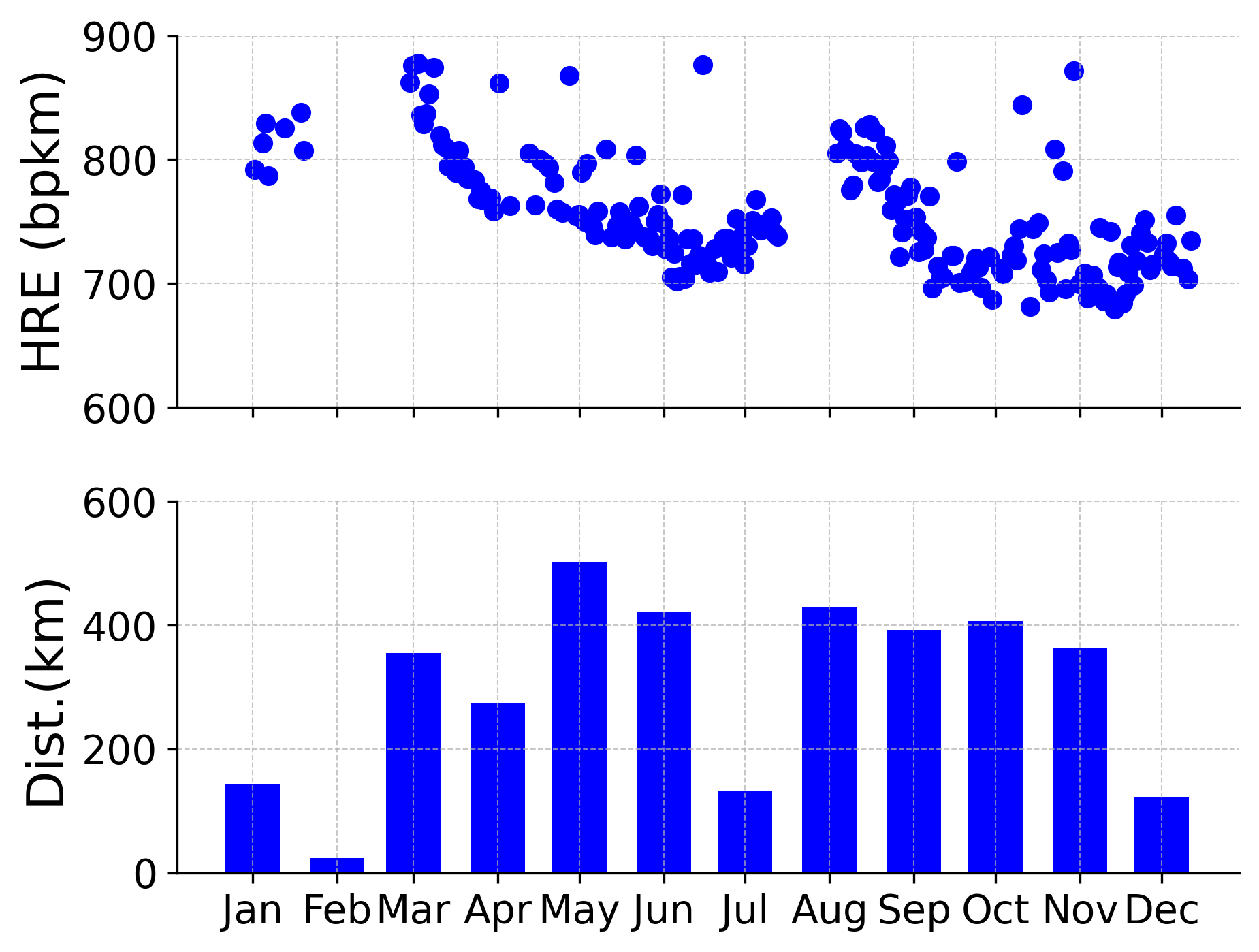}
        \caption{2014}
    \end{subfigure}

    \medskip

    \begin{subfigure}[b]{0.22\textwidth}
        \includegraphics[width=\textwidth]{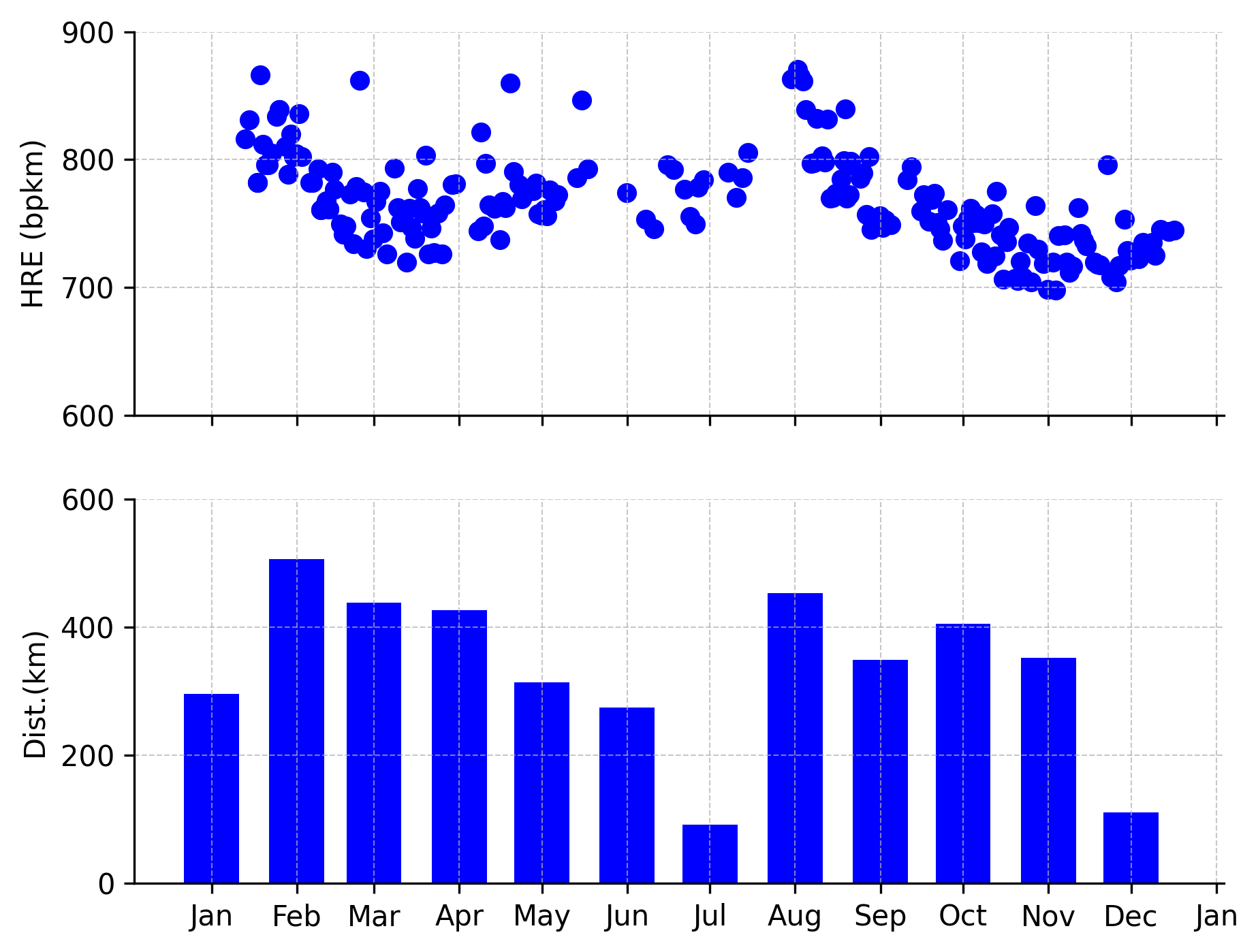}
        \caption{2015}
    \end{subfigure}
    \hfill
    \begin{subfigure}[b]{0.22\textwidth}
        \includegraphics[width=\textwidth]{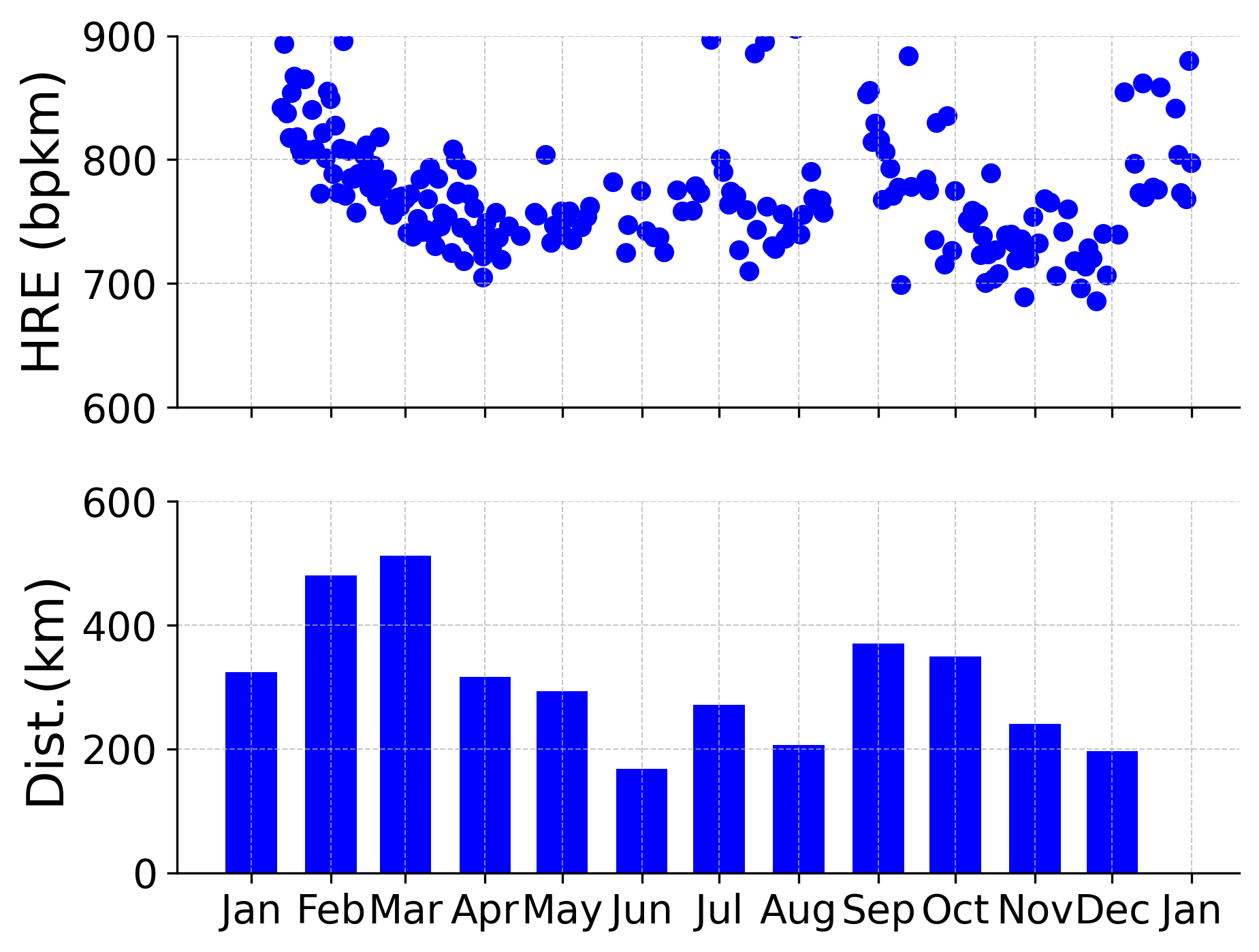}
        \caption{2016}
    \end{subfigure}
    \hfill
    \begin{subfigure}[b]{0.22\textwidth}
        \includegraphics[width=\textwidth]{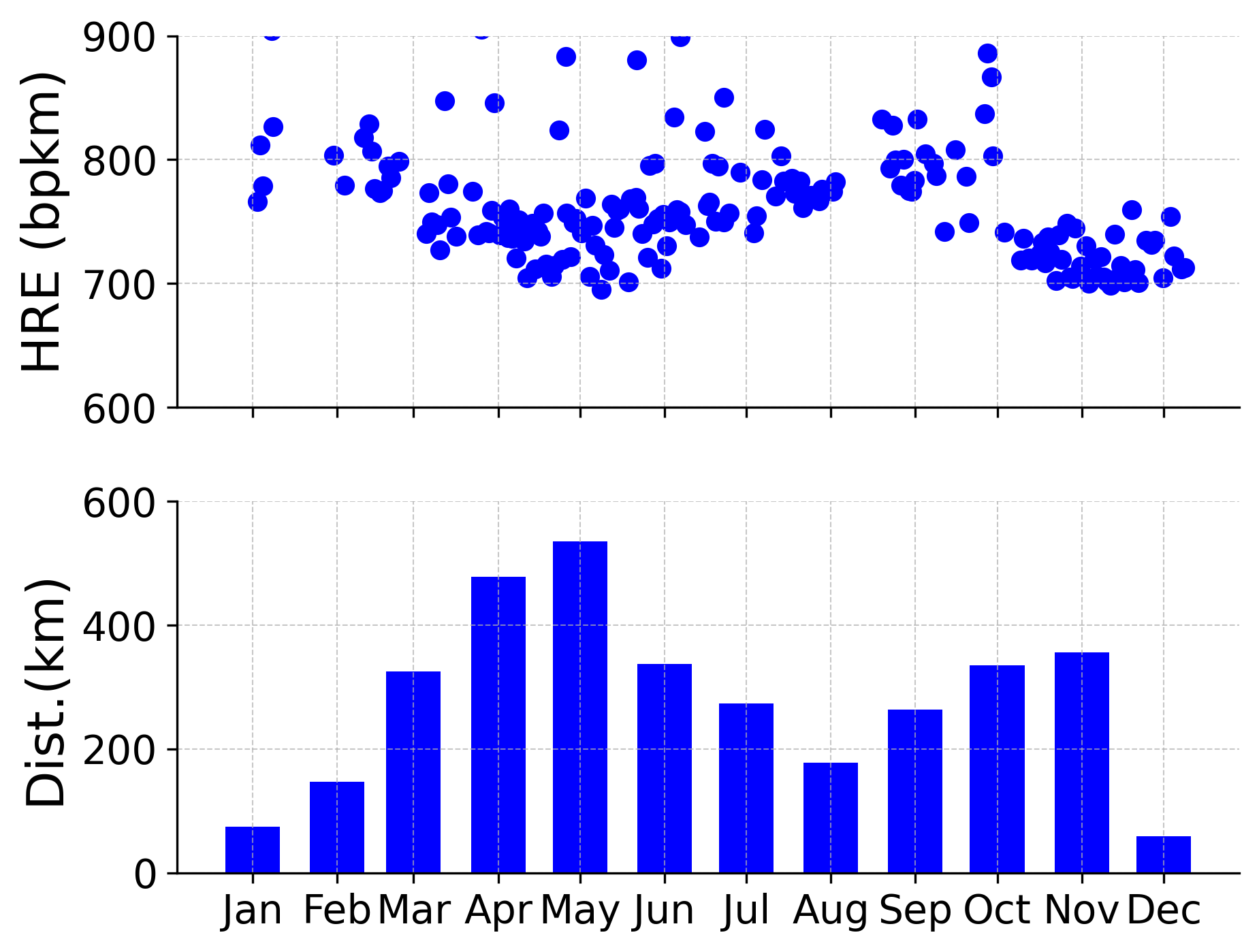}
        \caption{2017}
    \end{subfigure}
    \hfill
    \begin{subfigure}[b]{0.22\textwidth}
        \includegraphics[width=\textwidth]{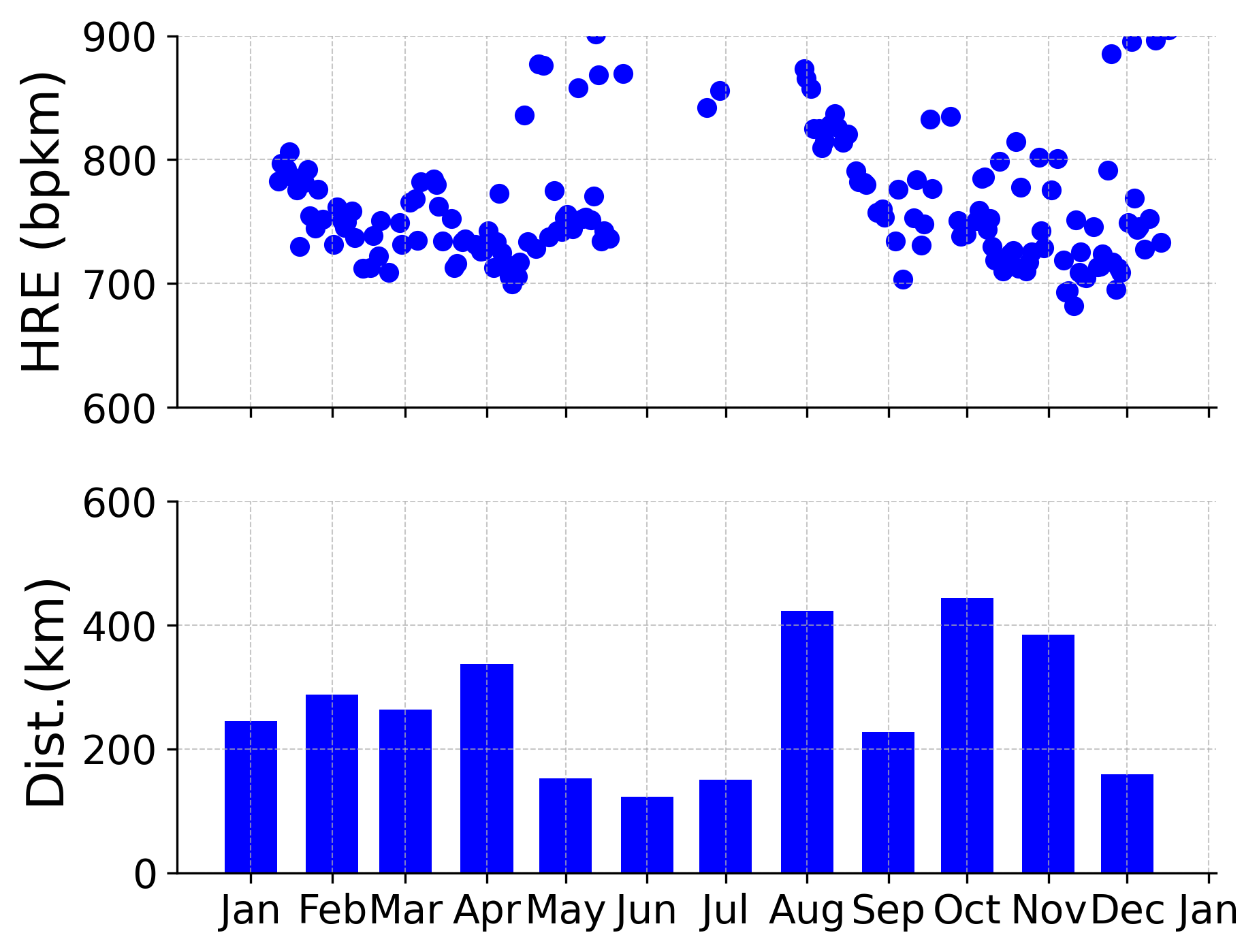}
        \caption{2018}
    \end{subfigure}

    \medskip

    \begin{subfigure}[b]{0.22\textwidth}
        \includegraphics[width=\textwidth]{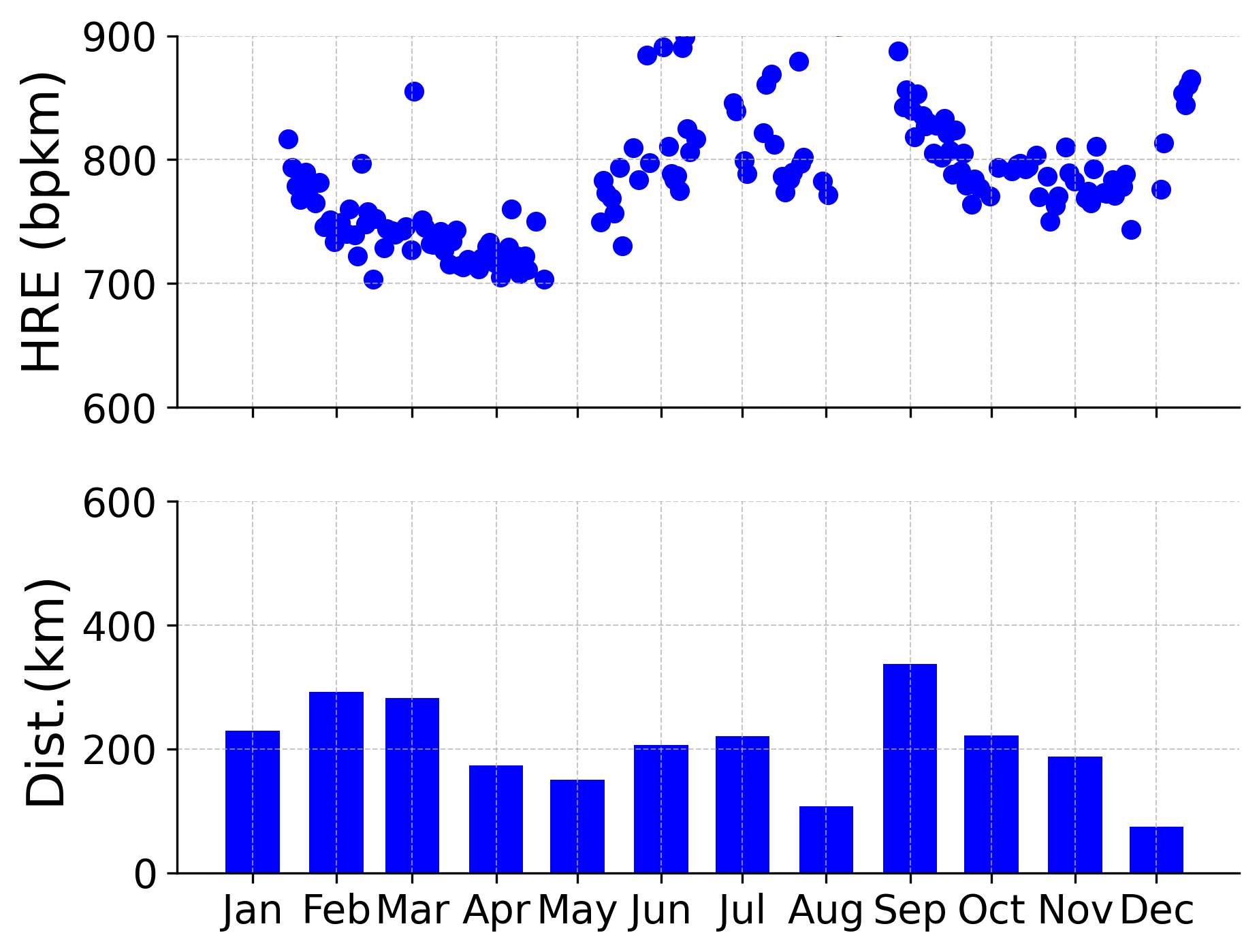}
        \caption{2019}
    \end{subfigure}
    \hfill
    \begin{subfigure}[b]{0.22\textwidth}
        \includegraphics[width=\textwidth]{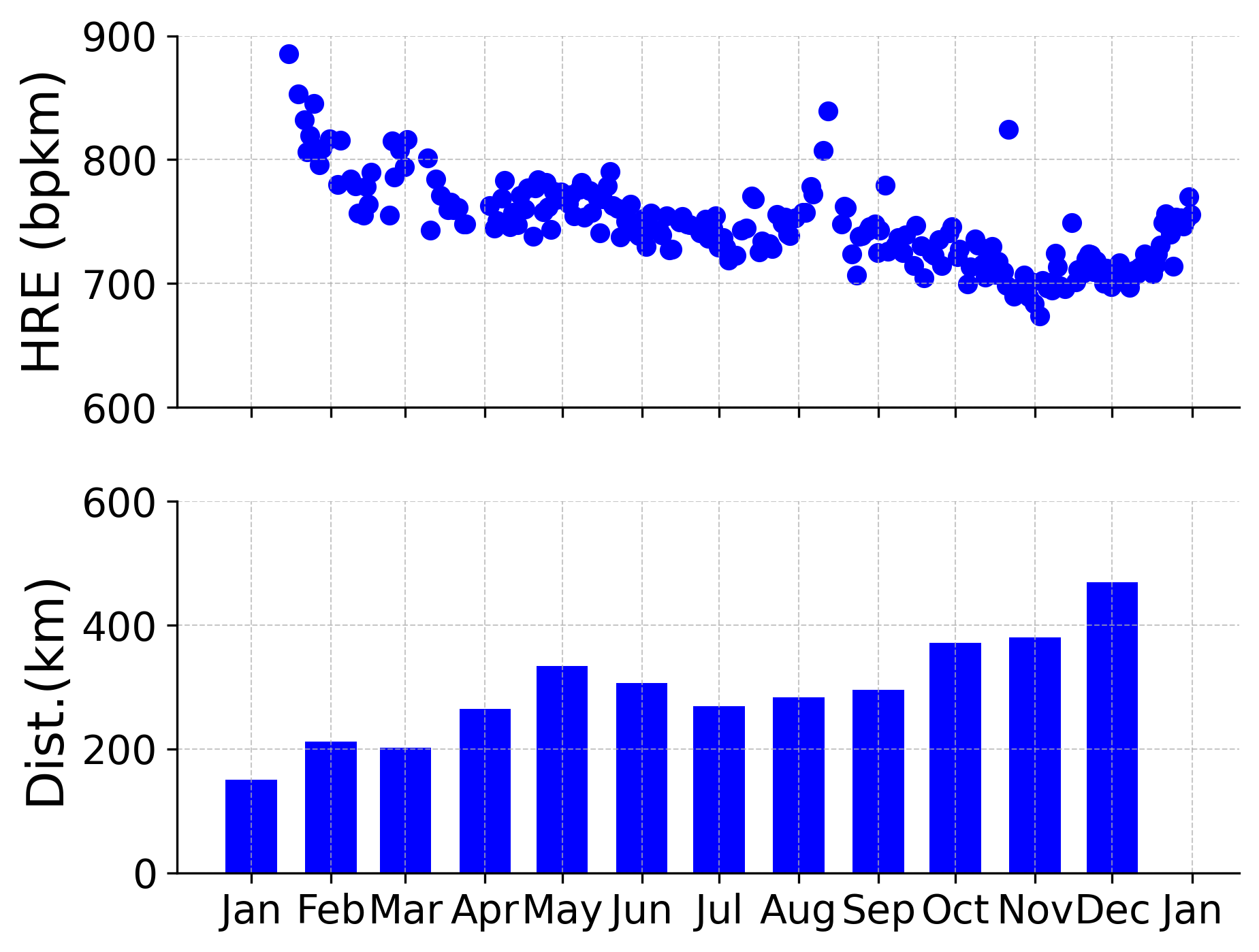}
        \caption{2020}
    \end{subfigure}
    \hfill
    \begin{subfigure}[b]{0.22\textwidth}
        \includegraphics[width=\textwidth]{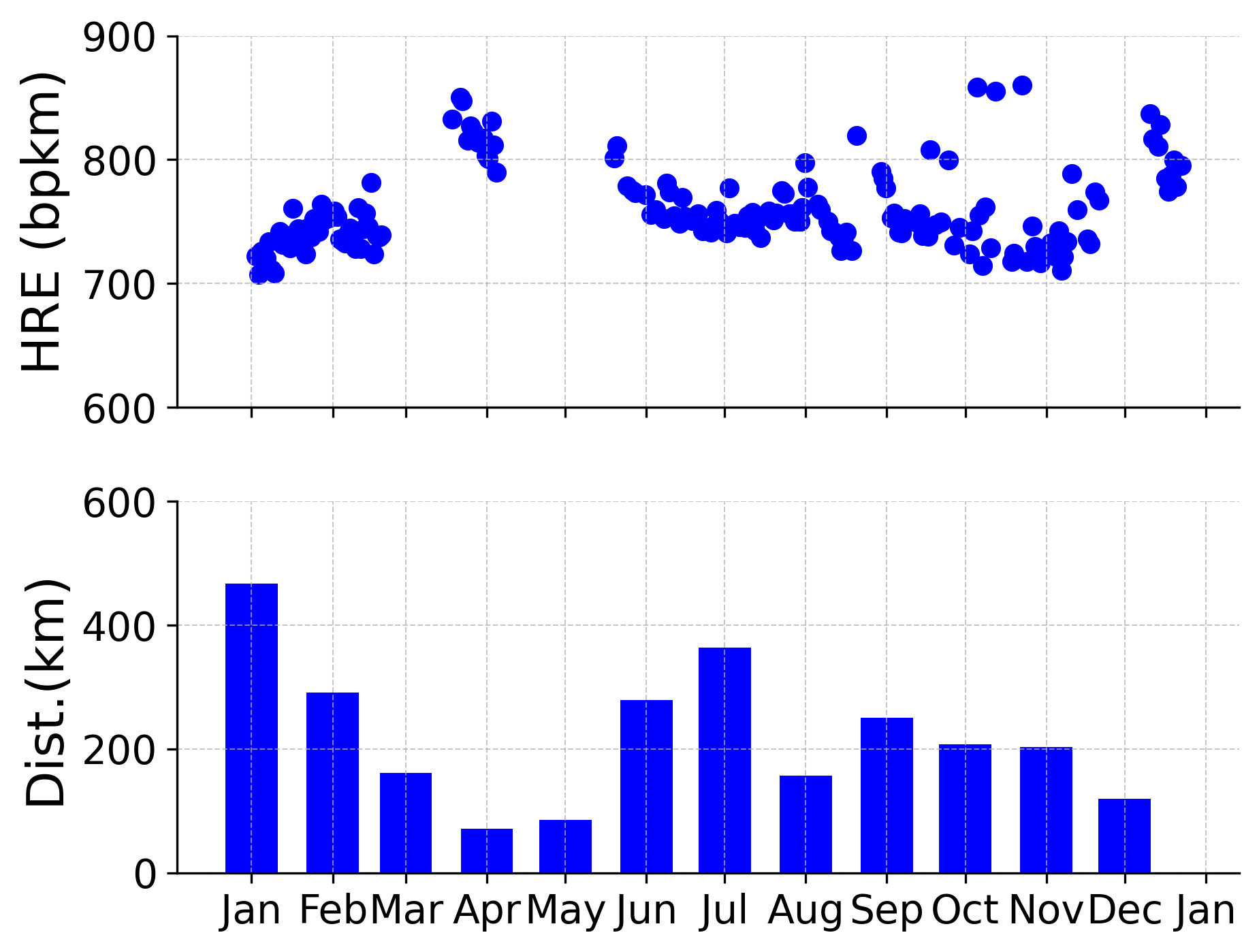}
        \caption{2021}
    \end{subfigure}
    \hfill
    \begin{subfigure}[b]{0.22\textwidth}
        \includegraphics[width=\textwidth]{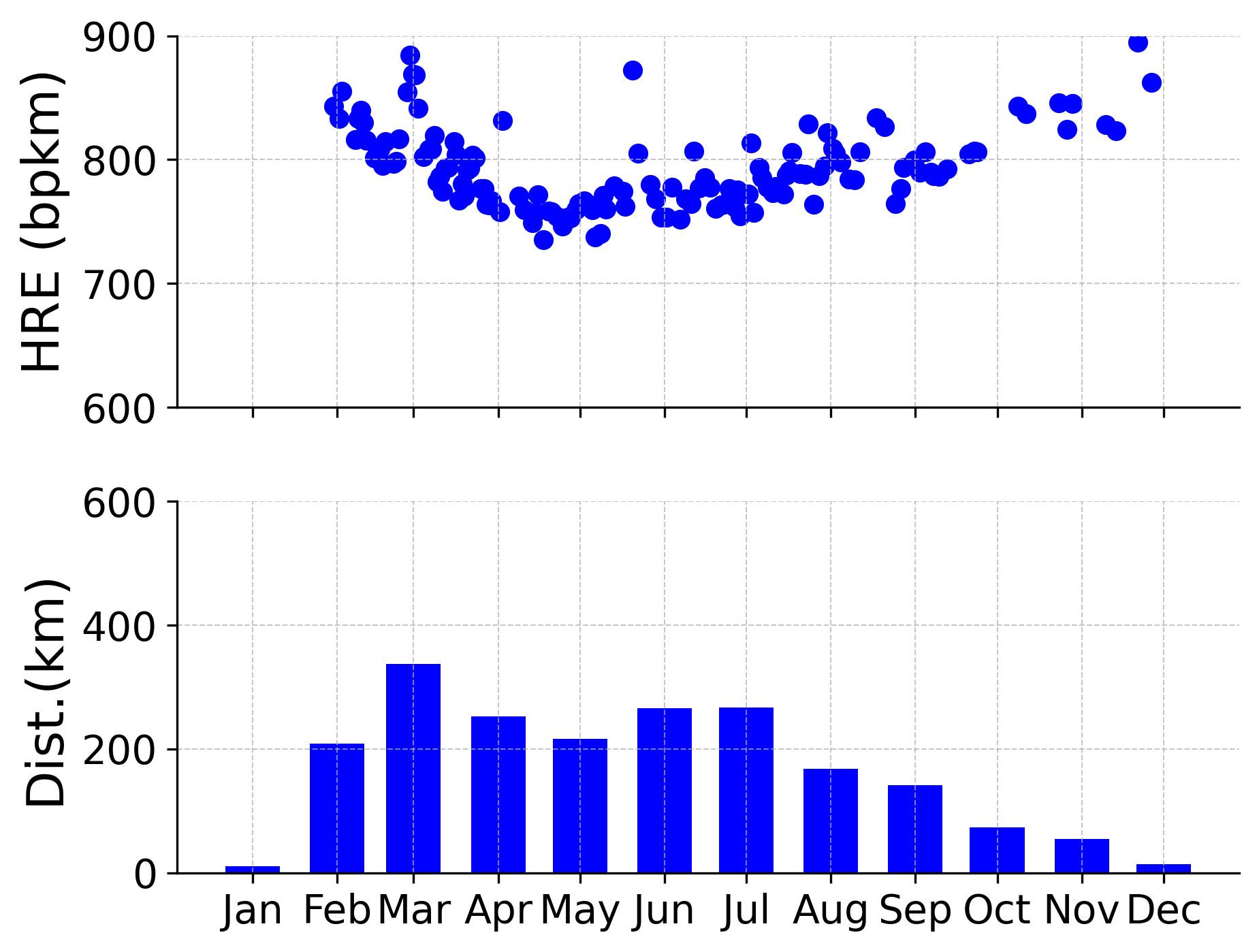}
        \caption{2022}
    \end{subfigure}

    \caption{HRE evolution and monthly distance for Athlete A from 2011 to 2022. Each subplot shows individual runs (markers) over time. HRE trends downward during periods of consistent training (2011–2016) and regresses slightly in later years.}
    \label{fig:HRE_years}
\end{figure*}

\begin{figure*}[t!]
    \centering
    \begin{subfigure}[b]{0.22\textwidth}
        \includegraphics[width=\textwidth]{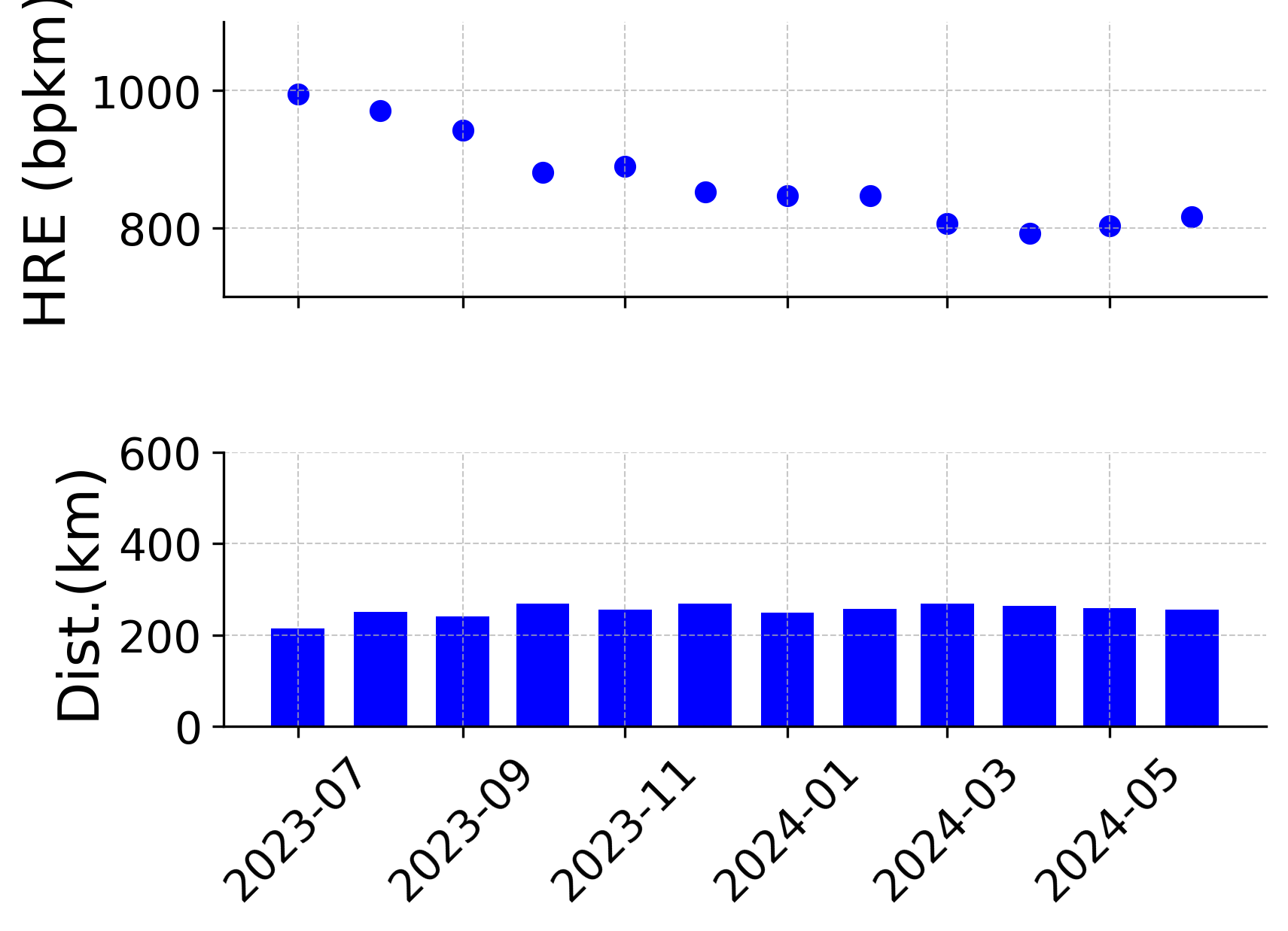}
        \caption{Athlete B}
    \end{subfigure}
    \hfill
    \begin{subfigure}[b]{0.22\textwidth}
        \includegraphics[width=\textwidth]{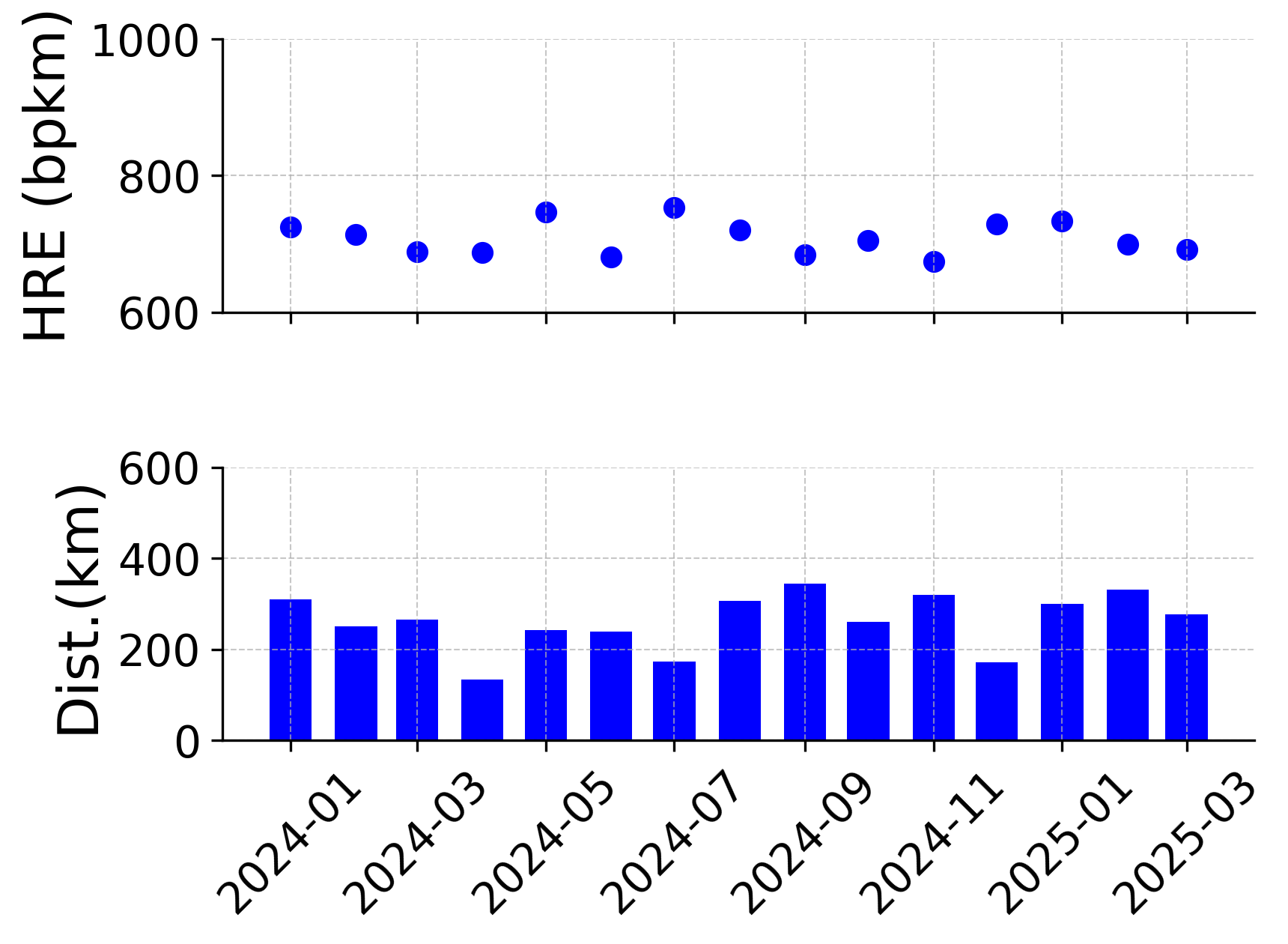}
        \caption{Athlete C}
    \end{subfigure}
    \hfill
    \begin{subfigure}[b]{0.22\textwidth}
        \includegraphics[width=\textwidth]{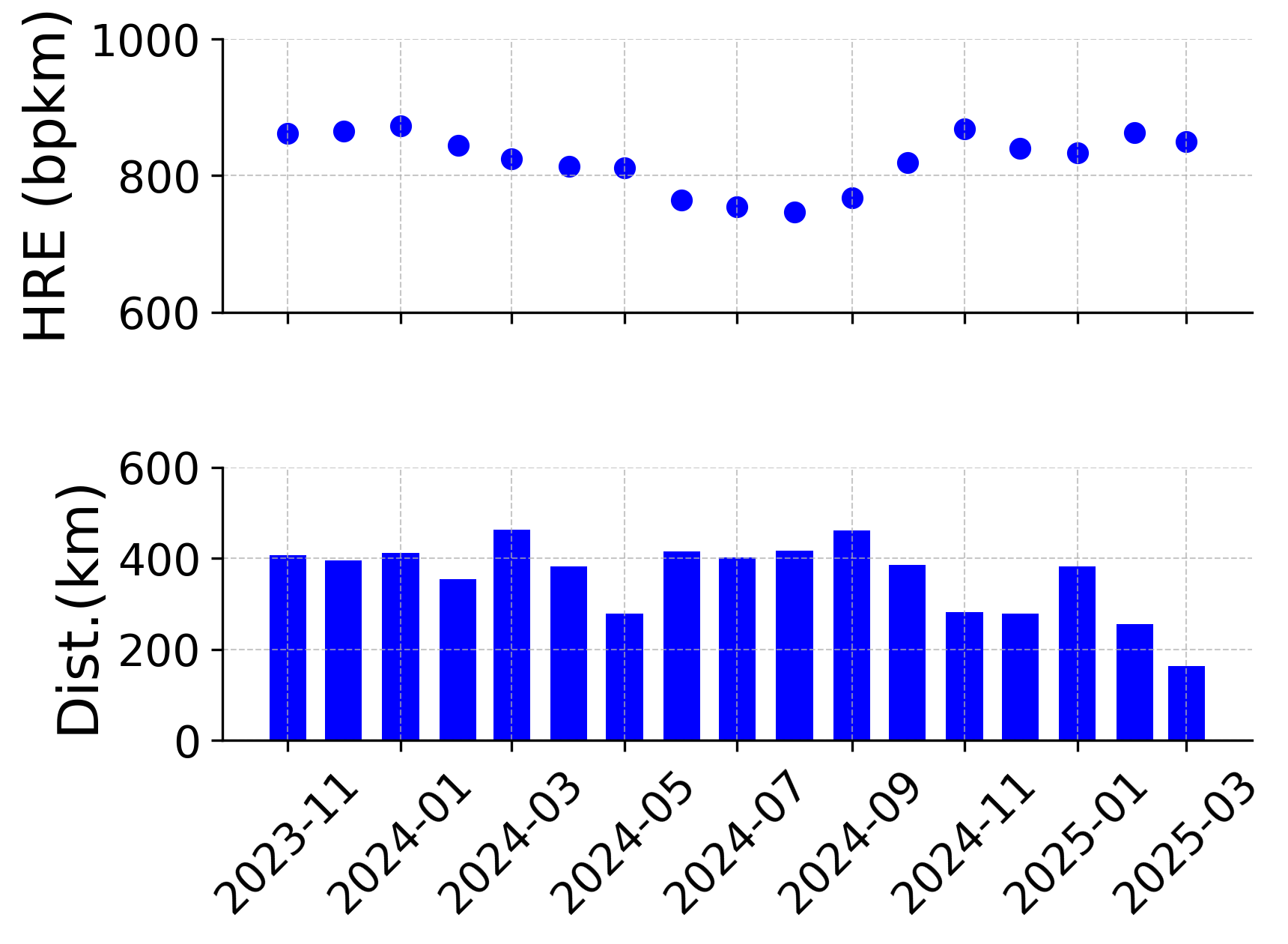}
        \caption{Athlete D}
    \end{subfigure}
    \hfill
    \begin{subfigure}[b]{0.22\textwidth}
        \includegraphics[width=\textwidth]{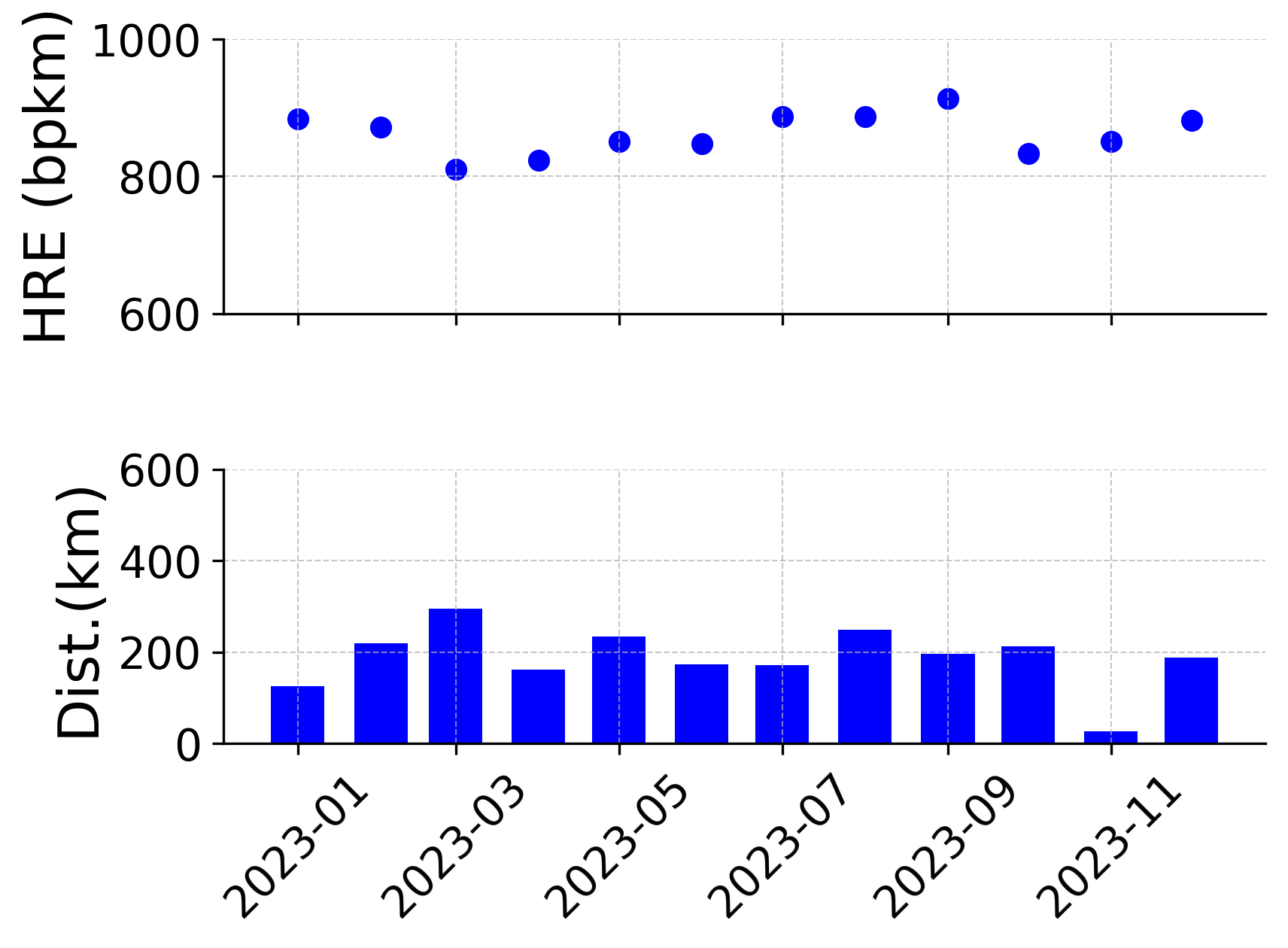}
        \caption{Athlete E}
    \end{subfigure}

    \medskip

    \begin{subfigure}[b]{0.22\textwidth}
        \includegraphics[width=\textwidth]{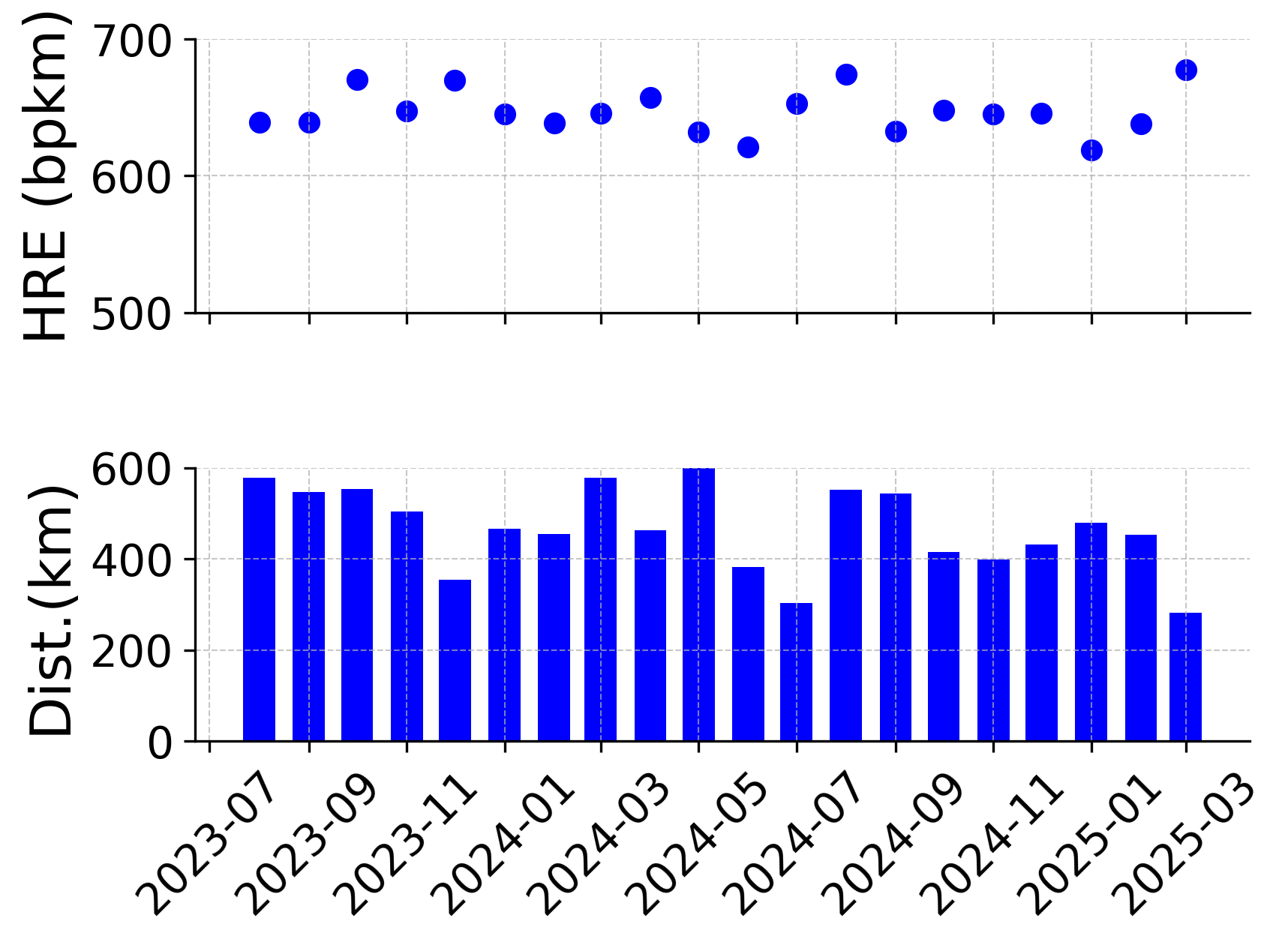}
        \caption{Athlete F}
    \end{subfigure}
    \hfill
    \begin{subfigure}[b]{0.22\textwidth}
        \includegraphics[width=\textwidth]{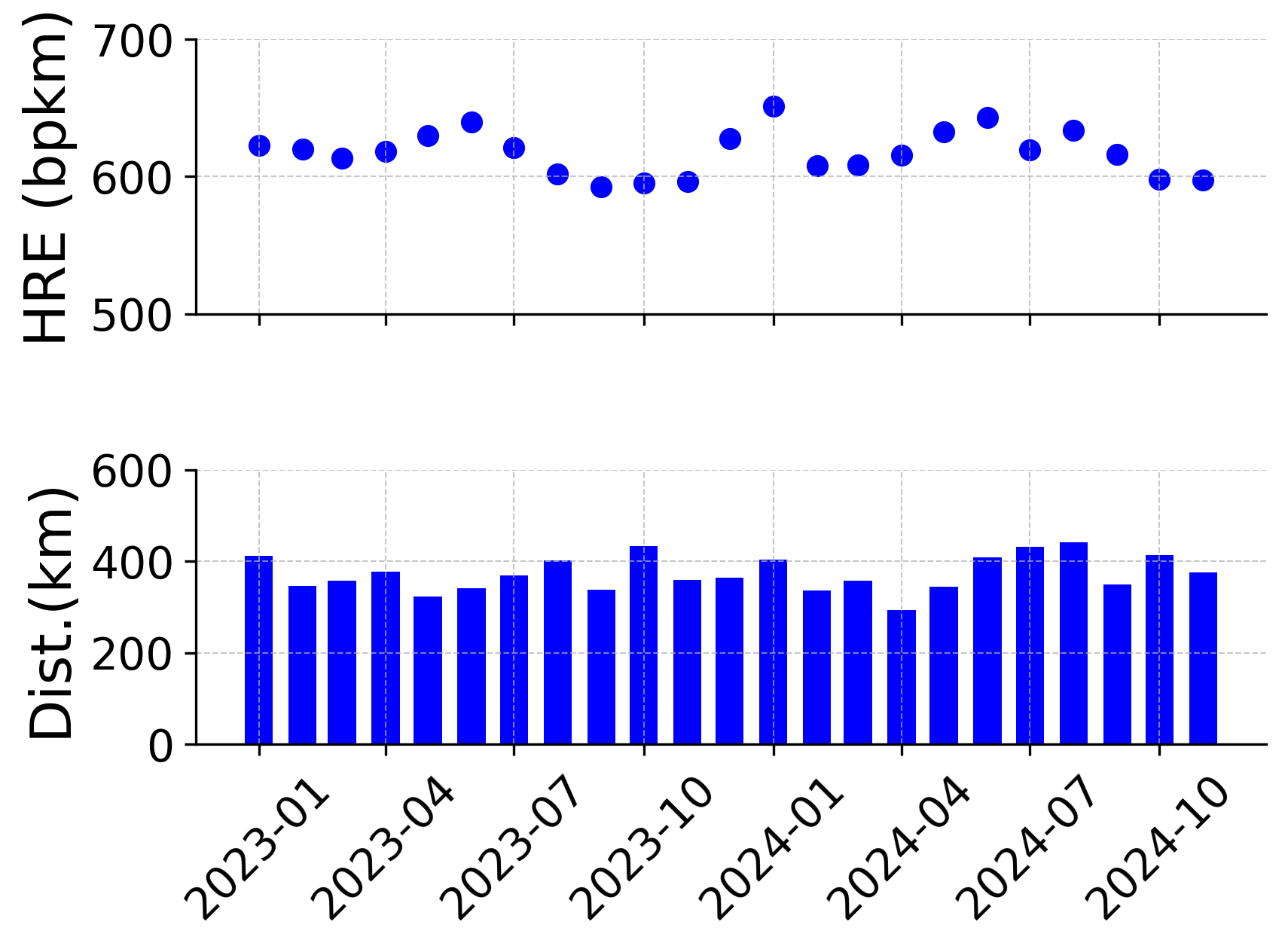}
        \caption{Athlete G}
    \hfill
    \end{subfigure}
    \begin{subfigure}[b]{0.22\textwidth}
        \includegraphics[width=\textwidth]{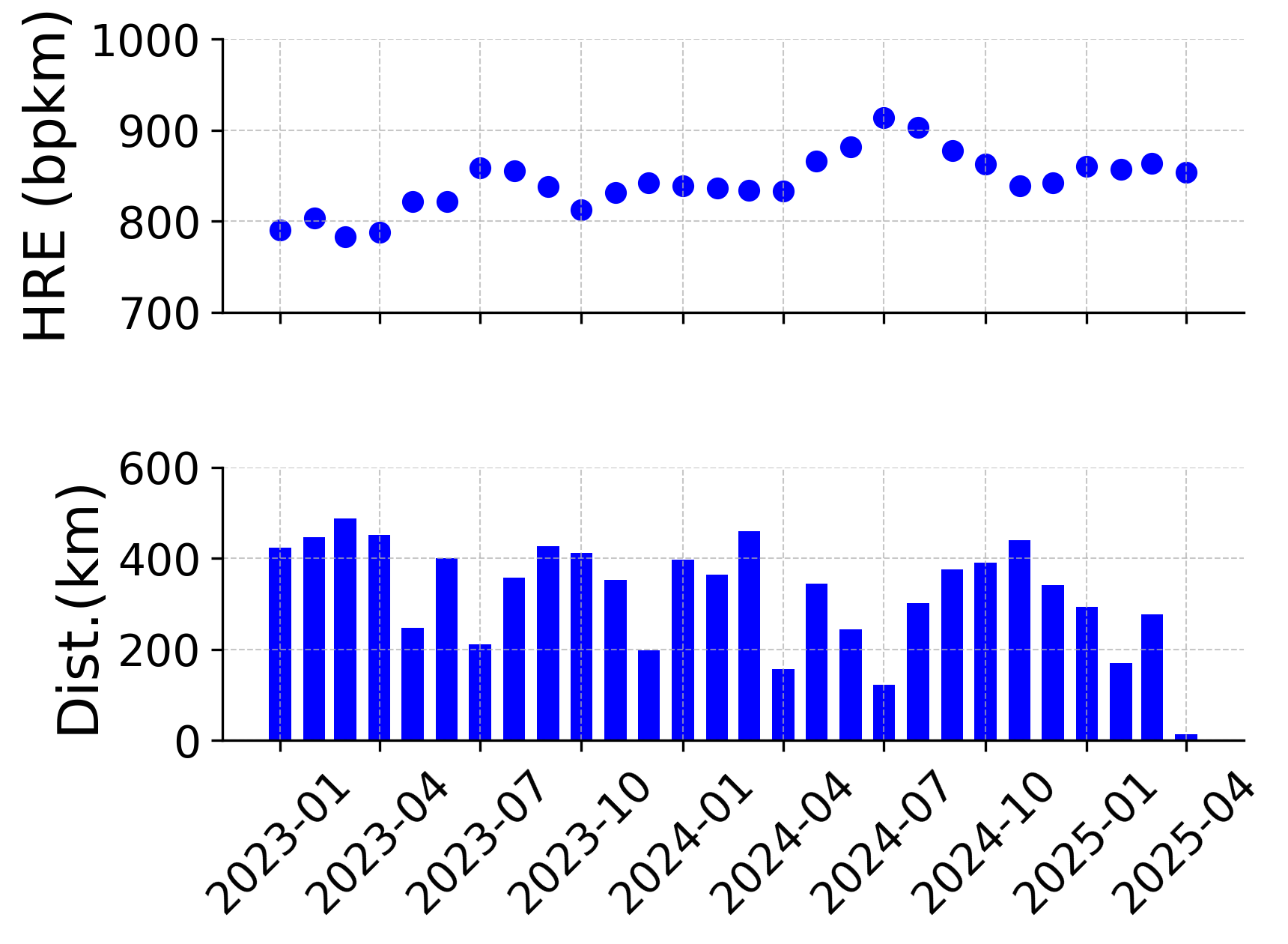}
        \caption{Athlete H}
    \end{subfigure}
    \hfill
    \begin{subfigure}[b]{0.22\textwidth}
        \includegraphics[width=\textwidth]{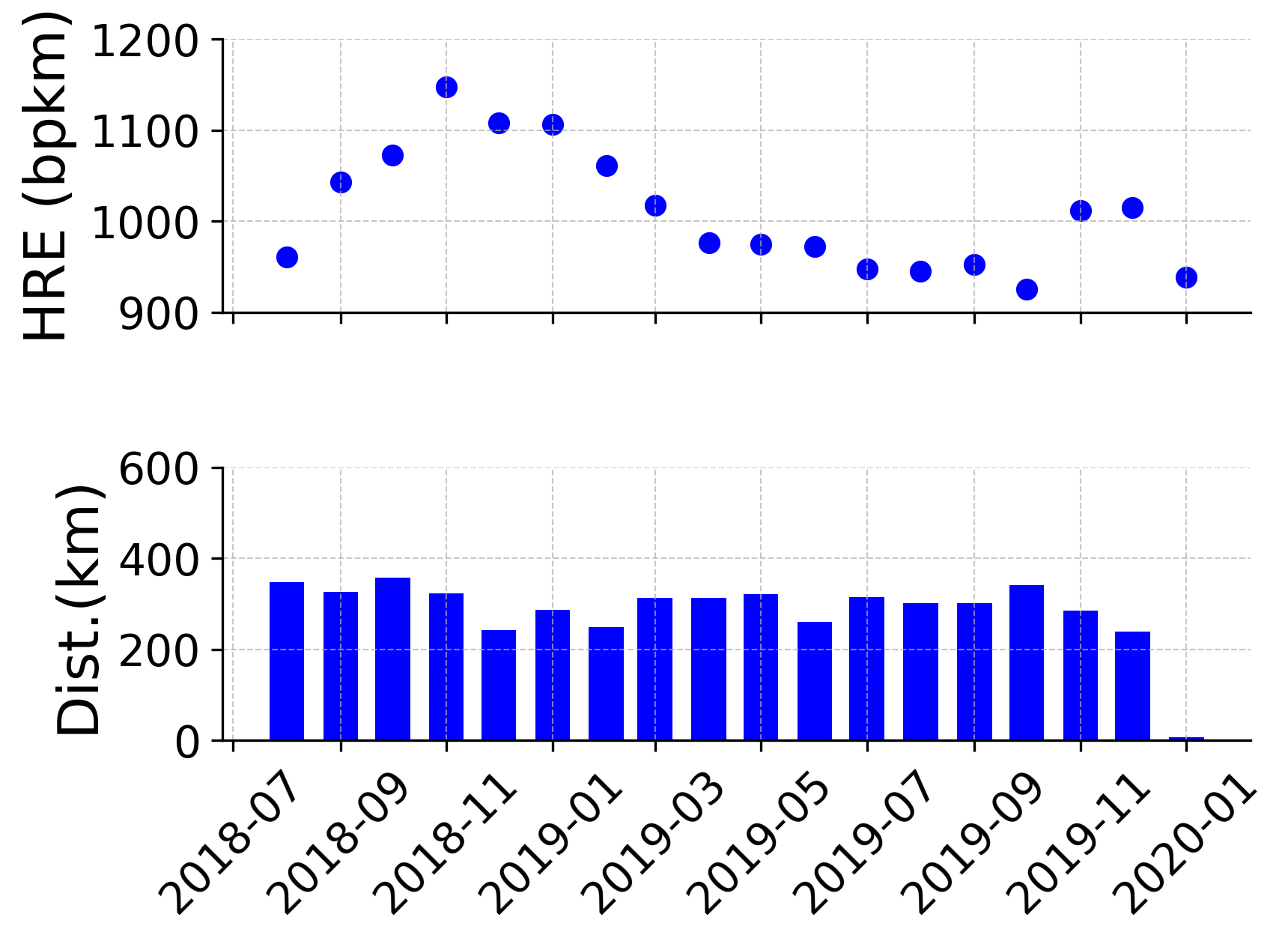}
        \caption{Athlete I}
    \end{subfigure}

    \medskip

    \begin{subfigure}[b]{0.22\textwidth}
        \includegraphics[width=\textwidth]{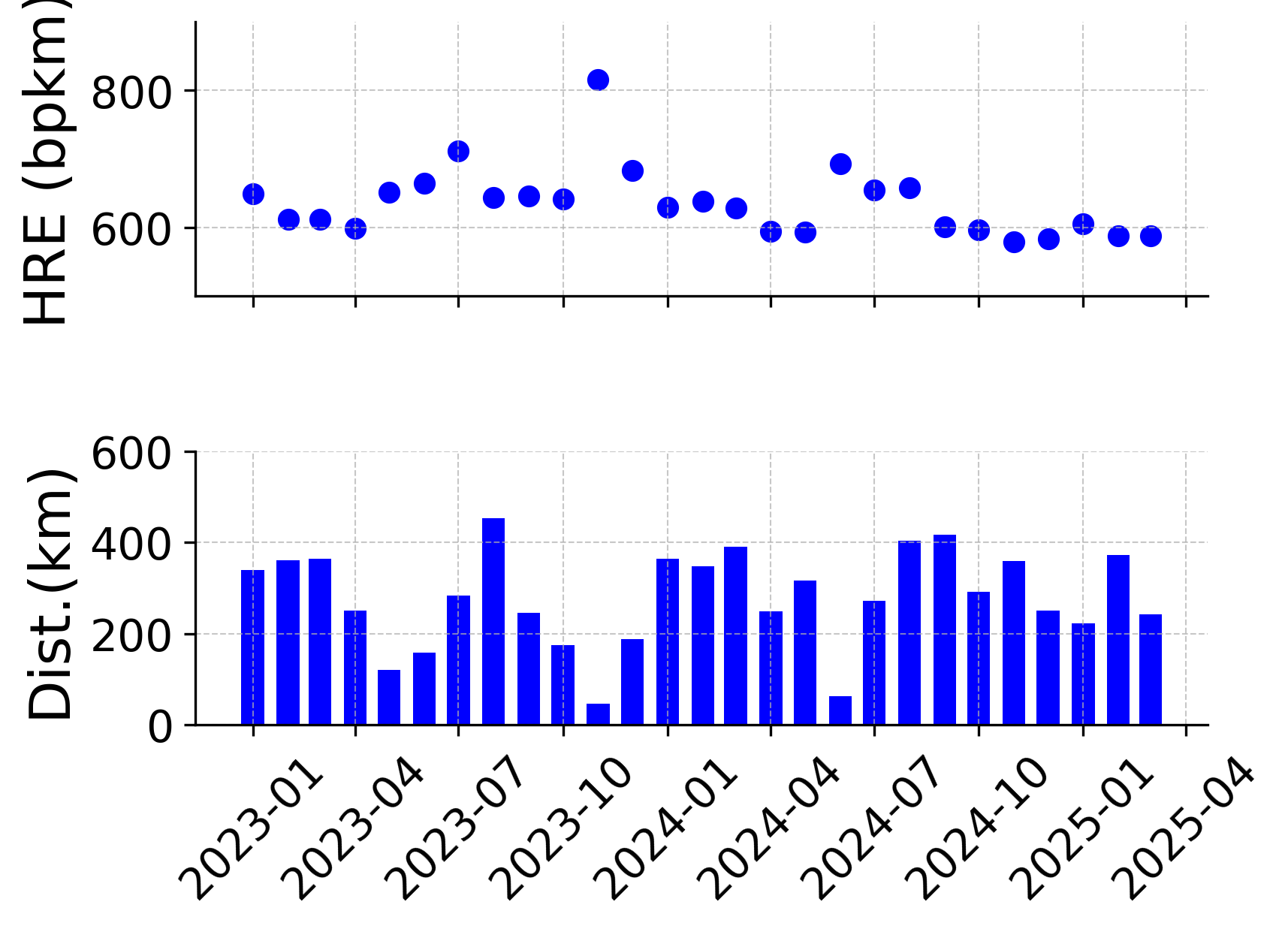}
        \caption{Athlete J}
    \end{subfigure}
    \hfill
    \begin{subfigure}[b]{0.22\textwidth}
        \includegraphics[width=\textwidth]{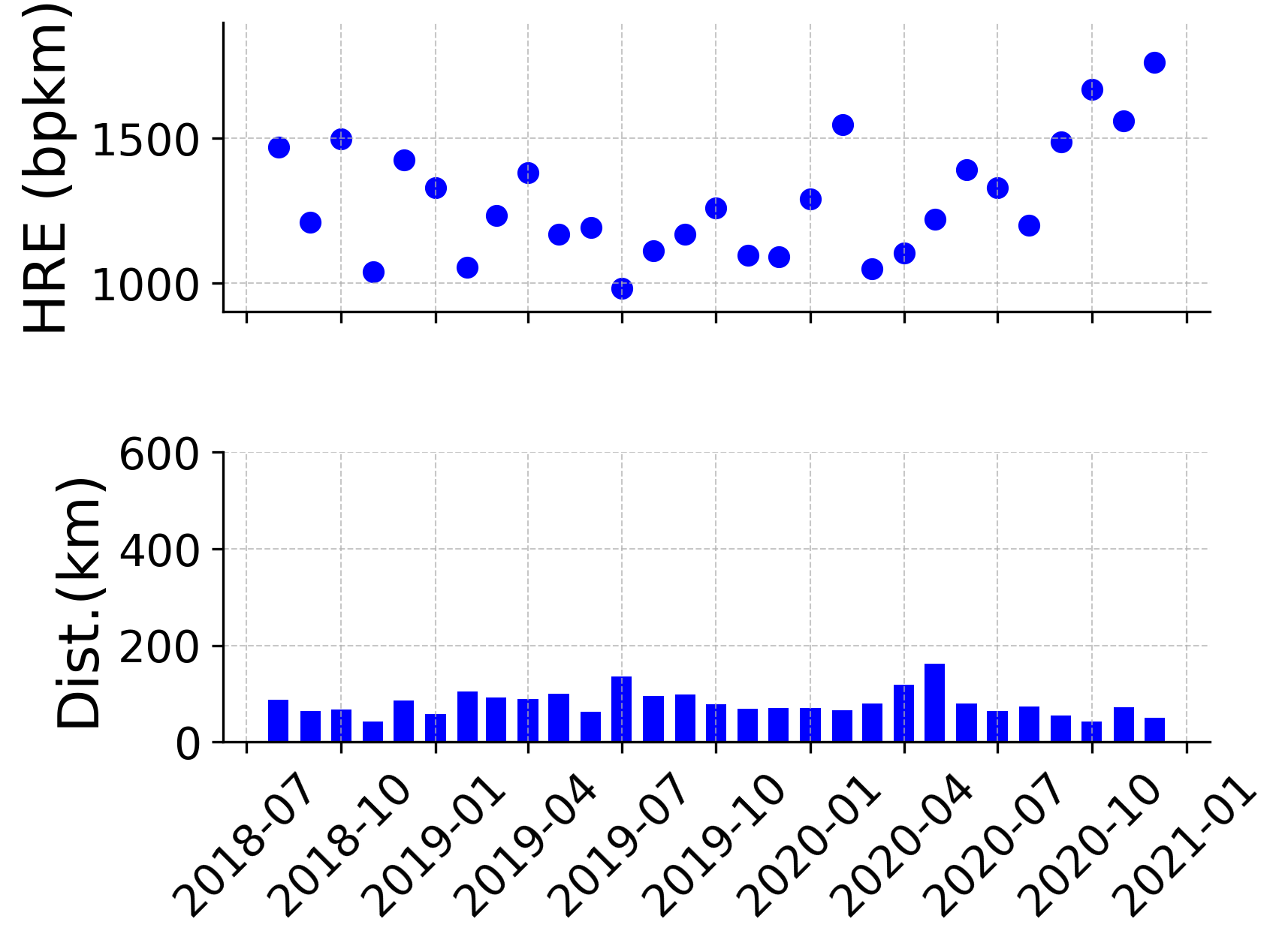}
        \caption{Athlete K}
    \end{subfigure}
    \hfill
    \begin{subfigure}[b]{0.22\textwidth}
        \includegraphics[width=\textwidth]{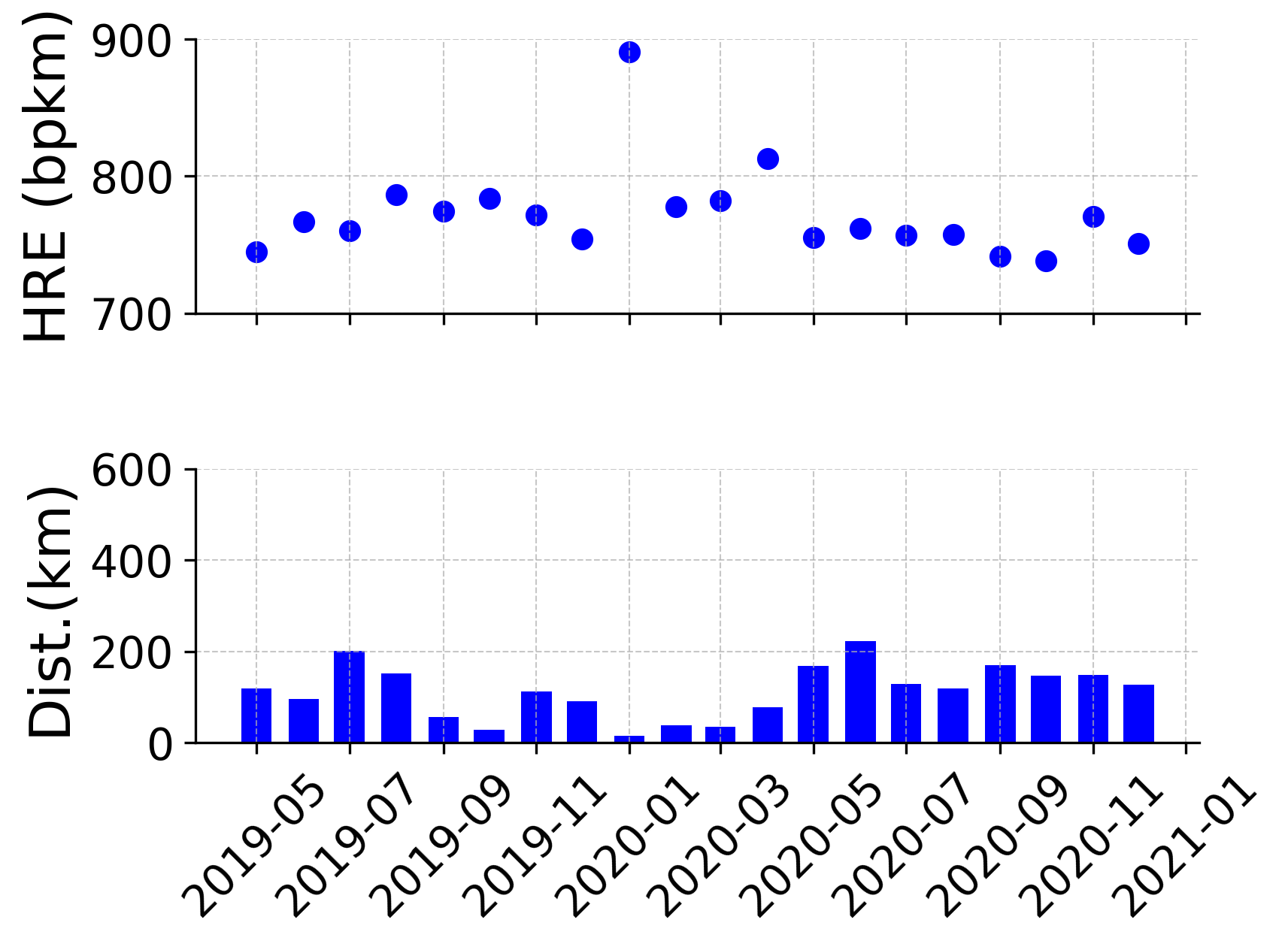}
        \caption{Athlete L}
    \end{subfigure}
    \hfill
    \begin{subfigure}[b]{0.22\textwidth}
        \includegraphics[width=\textwidth]{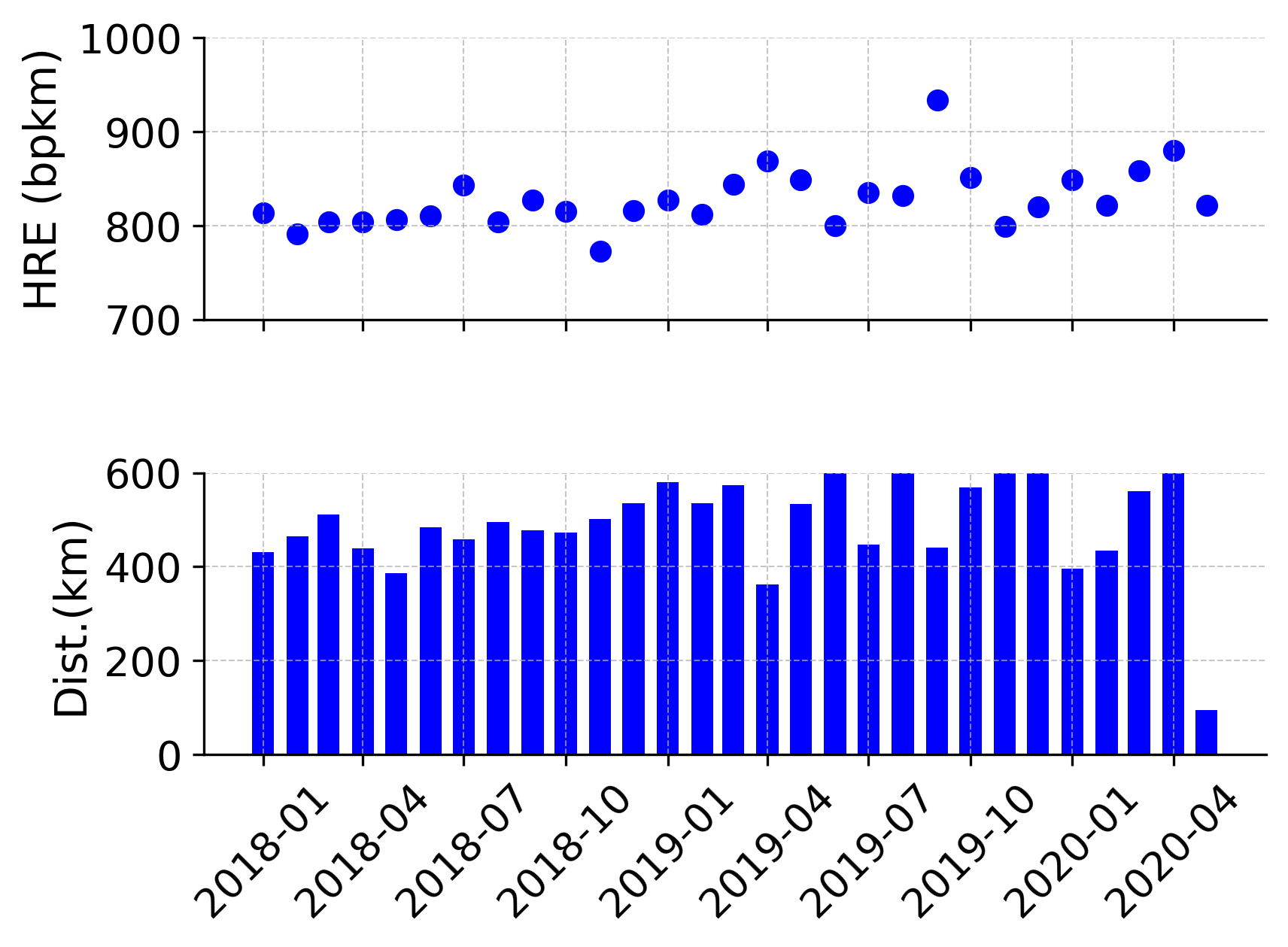}
        \caption{Athlete M}
        
    \end{subfigure}

    \caption{Monthly HRE and distance evolution across public runners. While experienced runners (F, G) show low and stable HRE, others (e.g., B, E) show seasonal variation (athlete K, low running volume, high HRE), reflecting differing training maturity and volume.}
    \label{fig:hre_public}
\end{figure*}

All analyses were conducted using Fitplotter, which enabled us to explore longitudinal and intra-run trends in HRE with synchronized visualizations and overlays.
\smallskip

\noindent\textbf{Longitudinal Evolution of Heart Rate Efficiency.} We analyzed 14 years of training data from athlete A. As shown in Figure~\ref{fig:HRE_years}, the athlete's HRE steadily improved during the initial three years of structured training (2011–2013), coinciding with increases in training volume. While average pace and heart rate fluctuated across seasons, HRE captured smoother trends and plateaus that aligned with performance peaks. Table~\ref{tab:yearly_summary}, summarizes the yearly averages for heart rate, pace, and HRE. A lower HRE reflects fewer heartbeats per kilometer, which is an indicator of improved cardiovascular economy. Athlete A's minimum HRE improved from 740 beats/km in 2011 (at age 45) to a personal best of 685 in 2016 (age 50, when he met the qualification standard for the Boston Marathon in the M50 category), suggesting primarily proper training organization, including the understanding and use of HRE in his training. In later years (2020–2024), a gradual decline in efficiency is observed, likely due to the natural effects of aging, reduced monthly mileage, and less frequent race participation. These findings suggest that HRE provides a stable and explainable signal for tracking long-term aerobic fitness progression, especially compared to pace or heart rate alone.
\smallskip

\noindent\textbf{Comparing Heart Rate Efficiency Across Runners.}
To assess generalizability, we analyzed 12 additional runners (B–M). Figure~\ref{fig:hre_public} presents the monthly HRE and training distance for each athlete. Runners F and G, both highly experienced and training at high volumes, showed low and stable HRE values (approx. 600–650 beats/km), whereas others (e.g., B, E) exhibited more seasonal fluctuations and higher HRE levels. These patterns support the use of HRE as a differentiator between levels of aerobic conditioning, independent of pace or raw mileage. For example, pace may vary due to temporary terrain changes or wind, and HR may spike due to caffeine or stress; HRE integrates these signals by offering a smoothed view of effort-per-distance.  Additionally, amateur runners often may misinterpret fitness metrics in isolation by focusing solely on achieving faster pace or keeping HR low. As observed in prior work~\cite{Smyth2022}, this can lead to counterproductive training decisions. In contrast, HRE encourages a more holistic view: regardless of the chosen pace or HR strategy, a decreasing HRE over time typically indicates improved aerobic fitness. This positions HRE as a more interpretable signal for non-experts navigating complex training data.
\smallskip

\begin{figure*}[t]
    \centering
    \begin{subfigure}[b]{0.49\textwidth}
        \includegraphics[width=0.9\textwidth]{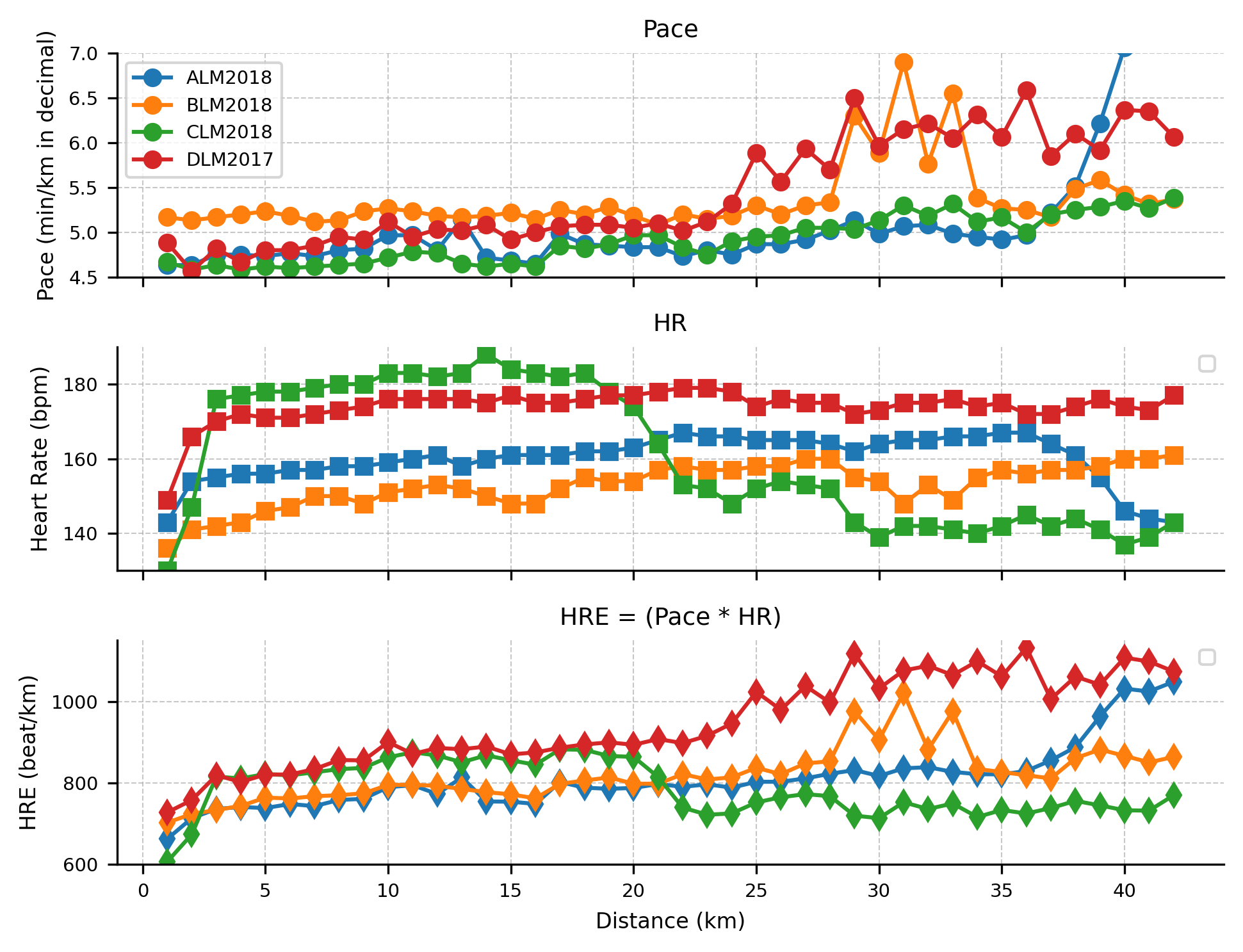}
        \caption{Poorly fitted runners}
    \end{subfigure}
    \hfill
    \begin{subfigure}[b]{0.49\textwidth}
        \includegraphics[width=0.9\textwidth]{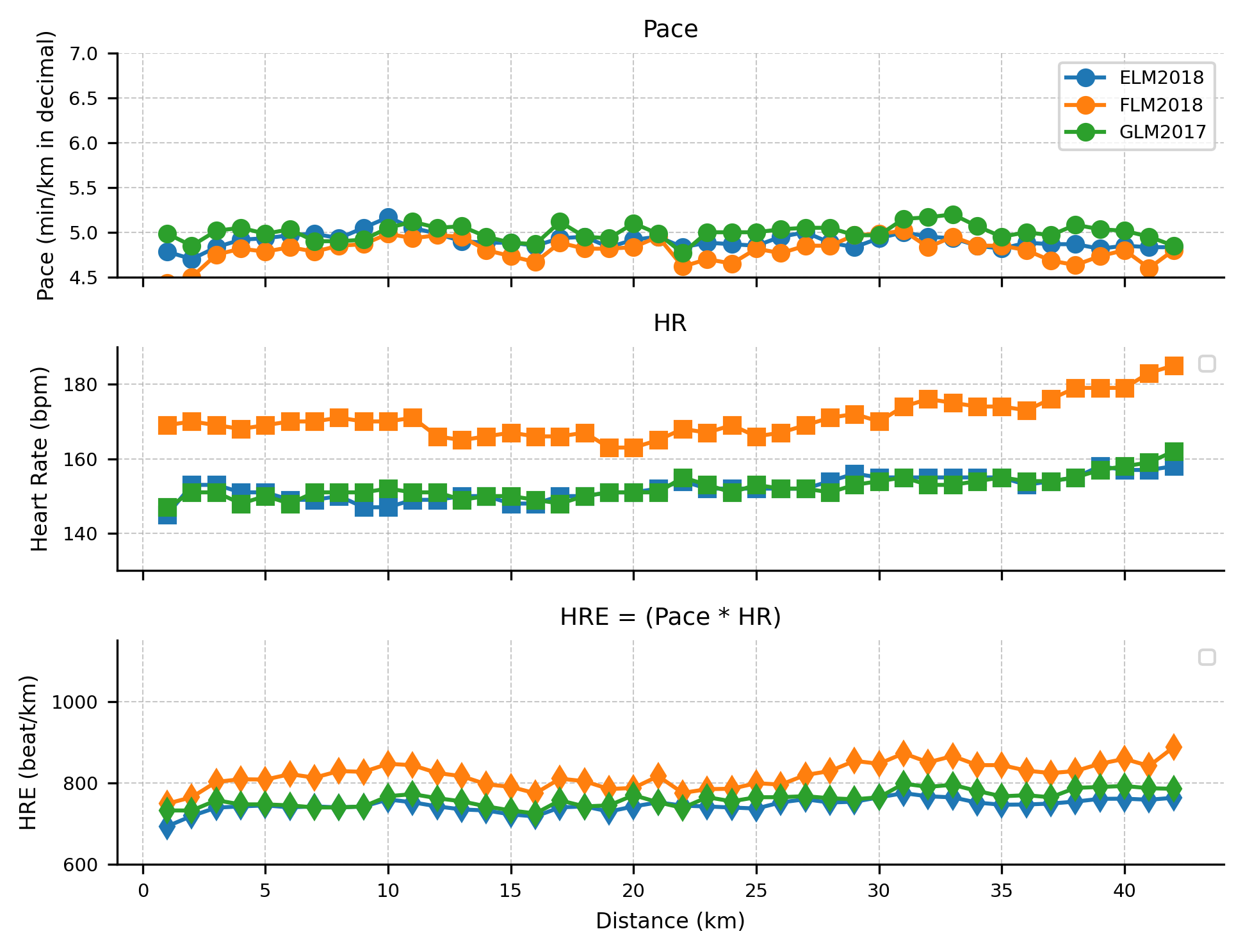}
        \caption{Well-fitted runners}
    \end{subfigure}
    \caption{Pace, HR, and HRE distributions of runners in the same race (Larnaca International Marathon 2017 and 2018). The poorly fitted runners (left) show  HRE degradation in the second half of the race, while the well-fitted runners (right) maintain stable HRE and consistent pacing. Empirical thresholds: HRE $<$ 700–750 indicates good fit; HRE $>$ 800 and degrading suggests poor fit.}
    \label{fig:HRE_LM}
\end{figure*}

\begin{figure*}[t]
    \centering
    \begin{subfigure}[b]{0.49\textwidth}
        \includegraphics[width=0.9\textwidth]{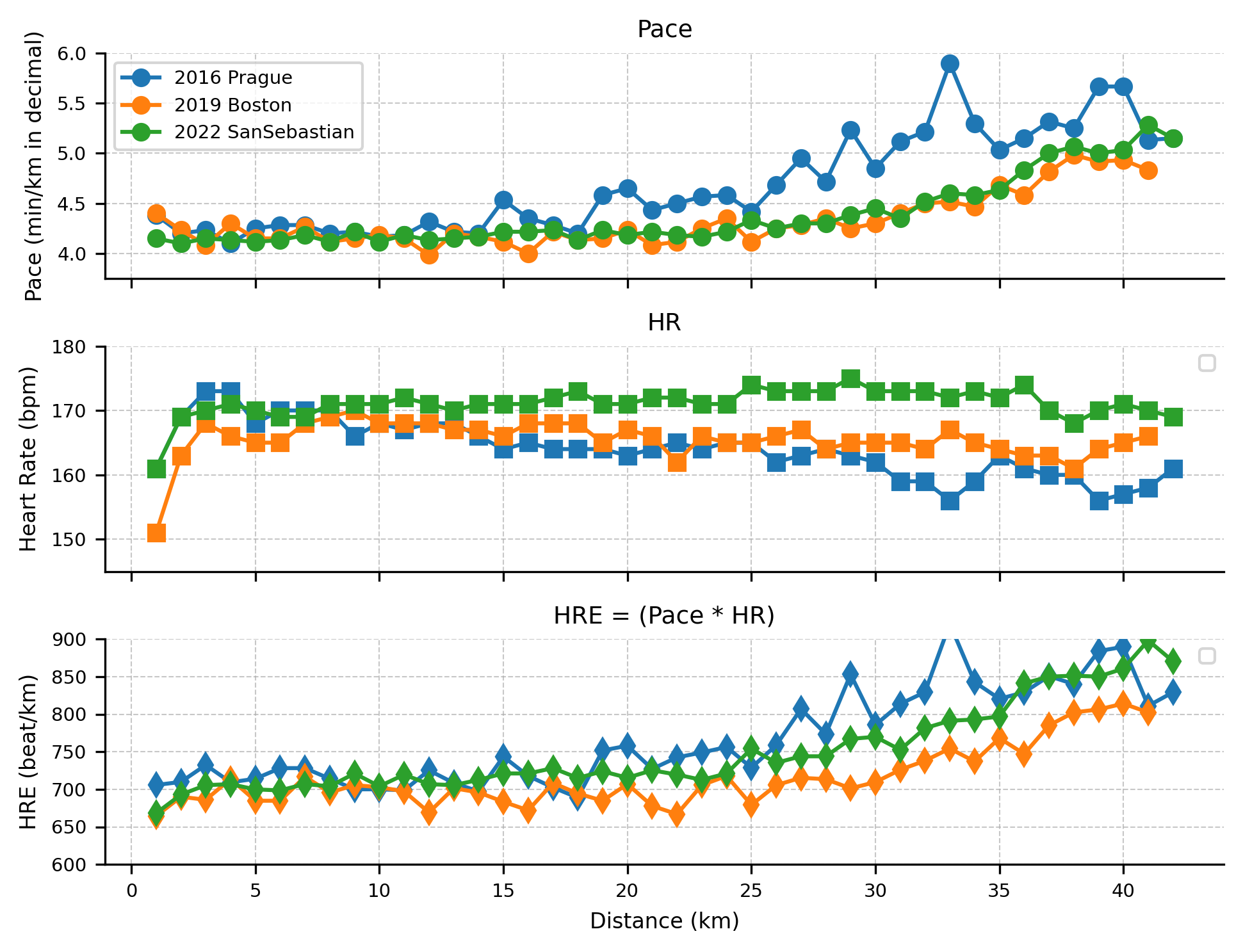}
        \caption{Lower fitness races}
    \end{subfigure}
    \hfill
    \begin{subfigure}[b]{0.49\textwidth}
        \includegraphics[width=0.9\textwidth]{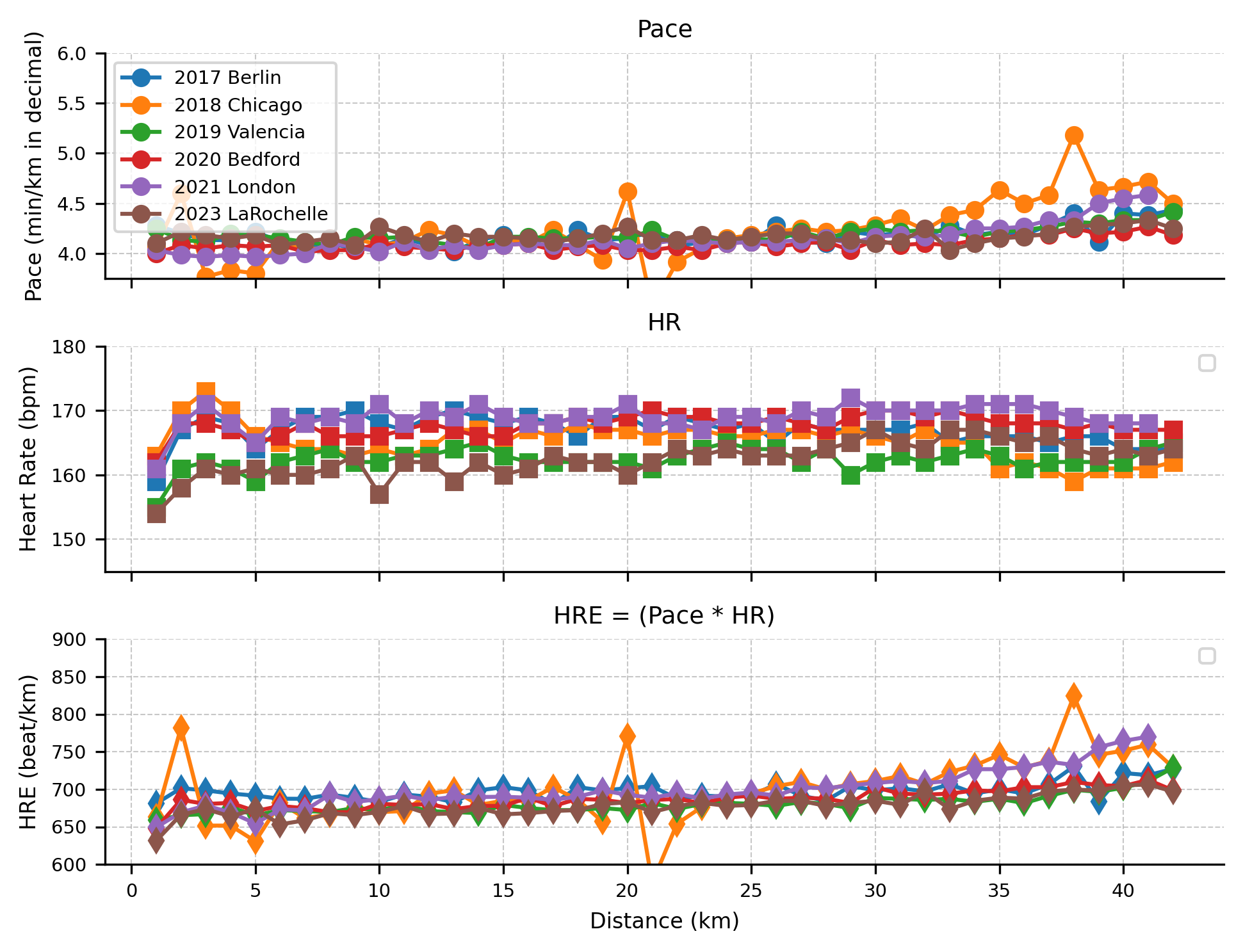}
        \caption{Higher fitness races}
    \end{subfigure}
    \caption{Comparison of pace, HR and HRE of the same runner in his  various marathons in nine sequent years. Right races demonstrates more consistent performance and lower HRE indicates improved fitness and pacing control.}
    \label{fig:HRE_NN}
\end{figure*}

\noindent\textbf{Interpreting HRE Around Competitive Events.} Finally, we analyzed HRE trends during competitive marathon events to investigate their relationship with peak performance. For the purposes of our analysis, we defined ``well-fitted'' runners as those having uniform HRE in the course of a long distance race. It correlates with HRE $<$ 700–750 beats/km for ``well-fitted'' and HRE $>$ 800 for ``poorly-fitted''. These cutoffs emerged from repeated observations across runners and events, and align with physiological expectations of efficiency. Figures~\ref{fig:HRE_LM} and~\ref{fig:HRE_NN} show HR, pace, and HRE data from two editions of the Larnaca Marathon (2017 and 2018), based on Strava activities from various runners whose activity IDs are listed in Table~\ref{tab:public_athletes}. Figures~\ref{fig:HRE_LM} and~\ref{fig:HRE_NN} also include data from major marathons run by the same athletes, all with sub-three-hour finishing times. Spikes in the curves are likely due to poor contact with the heart rate sensor.

The data are illustrated using two representative cases. In both cases, HRE remains steady for the first 30–35 km before deteriorating in the final kilometers; a pattern consistent with fatigue-induced breakdown (commonly referred to as ``hitting the wall''), as seen in the left panels (Figures~\ref{fig:HRE_LM} and~\ref{fig:HRE_NN}). In contrast, marathons with stable and typically lower HRE curves are associated with faster finishing times, as shown in the right panels (Figures~\ref{fig:HRE_LM} and~\ref{fig:HRE_NN}).

\section{Discussion}
\label{sec:discussion}

We revisited HRE as a simple and explainable metric for self-monitoring aerobic fitness in everyday training. Through longitudinal data analysis and comparative observations across multiple runners, we demonstrated that HRE can provide a meaningful signal of aerobic development than heart rate or pace alone. Beyond its empirical value, HRE also invites reflection on how fitness technologies can better support transparency, personalization, and user agency.
\smallskip

\noindent\textbf{Implications.} Our findings contribute to a growing body of research in HCI that calls for more explainable and user-centered fitness technologies. Prior work has shown that users often struggle to make sense of the metrics provided by commercial platforms, especially when those metrics are presented as black-box scores or algorithmically derived readiness indicators~\cite{epstein2015lived, choe2014understanding}. In contrast, HRE offers a transparent and contextually grounded measure that amateur runners can calculate and interpret on their own. This creates opportunities for self-tracking systems to support not only behavioral regulation but also a deeper understanding of personal physiological responses to training.

First, we suggest that designers consider integrating context-aware simplifications. As Epstein et al. \cite{epstein2015lived} argue, meaningful personal informatics systems should support users in interpreting data within their own routines and goals. Systems could automatically compute HRE in the background, surfacing it only when relevant changes occur (e.g., a period of consistent improvement, unexpected regression, or signs of physiological plateau). Unlike HR or pace, which can vary significantly even within the same session, HRE offers a synthesized and context-resistant view of performance. This is particularly useful for non-expert runners trying to make sense of weekly or seasonal training changes without overreacting to short-term noise. Second, tools should be designed to emphasize longitudinal insight over short-term feedback. Prior HCI research has shown that users often struggle to derive long-term meaning from daily or weekly fluctuations in metrics like pace or heart rate~\cite{rooksby2014personal}. By highlighting seasonal trends, recovery arcs, or training block comparisons, HRE-based visualizations can help users build a narrative of progression; this aligns with calls to design for personal meaning-making over pure optimization~\cite{baumer2015reflective, nunes2015self}. Third, systems can help users contextualize thresholds and plateaus. Rather than providing generic targets (e.g., ``you need to improve by 10\%''), fitness tools could surface personal bests or prior performance bands to help runners assess their current state in light of their own history. This echoes design goals from reflective informatics that advocate for systems that support introspection and self-comparison~\cite{baumer2015reflective}. Finally, we argue for designing interfaces that encourage reflection rather than perfection. As prior work on self-tracking has shown, emphasizing constant improvement can inadvertently lead to anxiety or disengagement~\cite{beer2017quantified}. By presenting HRE as a stable and personalized signal of fitness (not a competition or leaderboard) systems can support a healthier and more sustainable approach to training. This aligns with efforts to move beyond behaviorist framings of feedback toward more exploratory and empowering experiences~\cite{nunes2015self, choe2014understanding}.

Together, these implications point toward a design space where transparent, physiologically grounded metrics like HRE can act as anchors for interpretation, enabling amateur athletes to become more attuned to their own performance without needing expert intervention or proprietary systems. Tools such as Fitplotter show how privacy-preserving and transparent analytics can be made accessible for runners who wish to engage deeply with their training history.
\smallskip

\noindent\textbf{Limitations and Future Work.}
While our study draws on over a decade of consistent self-tracking and includes public logs from 12 additional runners, it remains exploratory in nature. HRE's generalizability across different populations (e.g., older adults, individuals with health conditions) or sports beyond running has yet to be evaluated. Additionally, contextual variables such as terrain or weather may hinder HRE interpretation, resulting in being unreasonably high.

Future research could explore how amateur athletes understand and use HRE feedback in situ, or compare HRE against lab-based physiological measures such as ${VO}_{2}$ max or lactate threshold. There is also potential to prototype mobile or wearable interfaces that incorporate HRE into training recommendations (e.g., highlighting when a runner is reaching a performance plateau, or suggesting recovery periods based on long-term HRE trends). More broadly, we see opportunities to extend HRE-like metrics into coaching interfaces, collaborative goal-setting tools, or reflective visualization systems designed to help users make sense of their training data over time.

\section{Conclusion}
\label{sec:conclusion}
We revisited HRE as a simple and explainable metric for monitoring aerobic fitness in amateur running. Through the analysis of over a decade of self-tracking data and comparative insights from public training logs, we showed that HRE offers a more stable and meaningful signal than heart rate or pace alone. Beyond its empirical validation, HRE invites new ways of designing feedback in fitness technologies by shifting the focus from black-box scores to transparent and user-understandable metrics. As wearables continue to shape how people engage with physical activity, tools such as HRE have the potential to support more reflective, personalized, and empowering training experiences.

\begin{acks}
The work has received funding from the European Union’s Horizon 2020 Research and Innovation Programme Grant Agreement No. 739578 and the Government of the Republic of Cyprus through the Deputy Ministry of Research, Innovation and Digital Policy. It has also received funding from the European Union under grant agreement No. 101135209. Views and opinions expressed are those of the authors and do not necessarily reflect those of the European Union. Neither the European Union nor the granting authority can be held responsible for them.
\end{acks}

\bibliographystyle{ACM-Reference-Format}
\bibliography{main}

\end{document}